\documentclass[chicago]{emulateapj_jw}

\usepackage{hyperref}
\usepackage{epsfig}
\usepackage{natbib}
\usepackage{graphicx}                
\usepackage{lscape}
\usepackage{verbatim}              
\usepackage{amsmath,amsthm,amsfonts,amsopn,amssymb} 
\usepackage{longtable}
\usepackage{pdflscape}
\usepackage{afterpage}
\usepackage{rotating}					
\usepackage{ulem}
\usepackage{epstopdf}  
\usepackage{multirow}
\usepackage{color}
\usepackage{graphicx}
\usepackage[left]{lineno}
\newcommand{\totalnum}{41~} 
\newcommand{\totalst}{38~} 
\newcommand{\totalsingle}{17~} 
\newcommand{\totaldouble}{14~} 
\newcommand{\totaltriple}{10~} 
\newcommand{\keckao}{30~} 
\newcommand{\roboao}{11~} 
\newcommand{\totalao}{33~} 
 
\newcommand{\cfopspec}{14~} 
\newcommand{\eaespec}{6~} 
 
\newcommand{\cfopspect}{8~} 

\received{}
\accepted{}

\shorttitle{Long-Period Exoplanets}
\shortauthors{Planet Hunters VIII}

\begin{document}
\title{Planet Hunters. VIII. Characterization of \totalnum Long-Period Exoplanet Candidates from {\it Kepler} Archival Data}
\author{
Ji Wang\altaffilmark{1,2},
Debra A. Fischer\altaffilmark{1},
Thomas Barclay\altaffilmark{3,4},
Alyssa Picard\altaffilmark{1},
Bo Ma\altaffilmark{5},
Brendan P. Bowler\altaffilmark{2,6},
Joseph R. Schmitt\altaffilmark{1},
Tabetha S. Boyajian\altaffilmark{1},
Kian J. Jek\altaffilmark{7},
Daryll LaCourse\altaffilmark{7},
Christoph Baranec\altaffilmark{8},
Reed Riddle\altaffilmark{2},
Nicholas M. Law\altaffilmark{9},
Chris Lintott\altaffilmark{10},
Kevin Schawinski\altaffilmark{11},
Dean Joseph Simister\altaffilmark{7},
Boscher Gr$\acute{e}$goire\altaffilmark{7},
Sean P. Babin\altaffilmark{7},
Trevor Poile\altaffilmark{7},
Thomas Lee Jacobs\altaffilmark{7},
Tony Jebson\altaffilmark{7},
Mark R. Omohundro\altaffilmark{7},
Hans Martin Schwengeler\altaffilmark{7},
Johann Sejpka\altaffilmark{7},
Ivan A.Terentev\altaffilmark{7},
Robert Gagliano\altaffilmark{7},
Jari-Pekka Paakkonen\altaffilmark{7},
Hans Kristian Otnes Berge\altaffilmark{7},
Troy Winarski\altaffilmark{7},
Gerald R. Green\altaffilmark{7},
Allan R. Schmitt\altaffilmark{7}
Martti H. Kristiansen\altaffilmark{7}
Abe Hoekstra\altaffilmark{7}
} 

\altaffiltext{0}{This publication has been made possible by the participation of more than 200,000 volunteers in the Planet Hunters project. Their contributions are individually acknowledged at http://www.planethunters.org/\#/acknowledgements}
\altaffiltext{1}{Department of Astronomy, Yale University, New Haven, CT 06511, USA}
\altaffiltext{2}{California Institute of Technology, 1200 East California Boulevard, Pasadena, CA 91101, USA}
\altaffiltext{3}{NASA Ames Research Center, M/S 244-30, Moffett Field, CA 94035, USA}
\altaffiltext{4}{Bay Area Environmental Research Institute, Inc., 560 Third Street West, Sonoma, CA 95476, USA}
\altaffiltext{5}{Department of Astronomy, University of Florida, 211 Bryant Space Science Center, Gainesville, FL 32611-2055, USA}
\altaffiltext{6}{Department of Astronomy, University of Texas at Austin, 2515 Speedway, Stop C1400, Austin, TX 78712}
\altaffiltext{7}{Planet Hunter}
\altaffiltext{8}{Institute for Astronomy, University of Hawai`i at M\={a}noa, Hilo, HI 96720-2700, USA}
\altaffiltext{9}{Department of Physics and Astronomy, University of North Carolina at Chapel Hill, Chapel Hill, NC 27599-3255, USA}
\altaffiltext{10}{Oxford Astrophysics, Denys Wilkinson Building, Keble Road, Oxford OX1 3RH}
\altaffiltext{11}{Institute for Astronomy, Department of Physics, ETH Zurich, Wolfgang-Pauli-Strasse 27, CH-8093 Zurich, Switzerland}

\begin{abstract}

The census of exoplanets is incomplete for orbital distances larger than 1 AU.  Here, we present \totalnum long-period planet candidates {{in \totalst systems}} identified by Planet Hunters based on {\it Kepler} archival data (Q0-Q17). Among them, \totalsingle exhibit only one transit, \totaldouble have two visible transits and \totaltriple have more than three visible transits. For planet candidates with only one visible transit, we estimate their orbital periods based on transit duration and host star properties. The majority of the planet candidates in this work (75\%) have orbital periods that correspond to distances of 1-3 AU from their host stars. We conduct follow-up imaging and spectroscopic observations to validate and characterize planet host stars. In total, we obtain adaptive optics images for \totalao stars to search for possible blending sources. Six stars have stellar companions within 4$^{\prime\prime}$. We obtain high-resolution spectra for \eaespec stars to determine their physical properties. Stellar properties for other stars are obtained from the NASA Exoplanet Archive and the Kepler Stellar Catalog by~\citet{Huber2014}. {{We validate 7 planet candidates that have planet confidence over 0.997 (3-$\sigma$ level). These validated planets include 3 single-transit planets (KIC-3558849b, KIC-5951458b, and KIC-8540376c), 3 planets with double transits (KIC-8540376b, KIC-9663113b, and KIC-10525077b), and 1 planet with 4 transits (KIC-5437945b). }}This work provides assessment regarding the existence of planets at wide separations {{and the associated false positive rate for transiting observation (17\%-33\%)}}. More than half of the long-period planets with at least three transits in this paper exhibit transit timing variations up to 41 hours, which suggest additional components that dynamically interact with the transiting planet candidates. The nature of these components can be determined by follow-up radial velocity and transit observations.  



\end{abstract}


\section{Introduction}
Since its launch in March of 2009, the NASA {\it Kepler} mission has been monitoring $\sim$160,000 stars in order to detect transiting extrasolar planets with high relative photometric precision~\citep[$\sim$20 ppm in 6.5 h,][]{Jenkins2010}. In May 2013, the {\it Kepler} main mission ended with the failure of a second reaction wheel; however, the first four years of {\it Kepler} data have led to a wealth of planetary discoveries with a total of 4,706 announced planet candidates\footnote{http://exoplanetarchive.ipac.caltech.edu/ as of Nov 11 2015}~\citep{Borucki2010,Borucki2011,Batalha2013,Burke2014}. The confirmed and candidate exoplanets typically have orbital periods shorter than 1000 days because at least three detected transits are needed for identification by the automated Transit Planet Search algorithm. Therefore, transiting exoplanets with periods longer than $\sim$1000 days are easily missed. The detection of short-period planets is further favored because the transit probability decreases linearly with increasing orbital distance. For these reasons, estimates of the statistical occurrence rate of exoplanets tend to focus on orbital periods shorter than a few hundred days \citep[e.g.,][]{Fressin2013,Petigura2013,Dong2013}.  Radial velocity (RV) techniques also favor the detection of shorter period orbits. While gas giant planets have been discovered with orbital periods longer than a decade, their smaller reflex velocity restricts detection of sub-Neptune mass planets to orbital radii less than $\sim$1 AU~\citep{Lovis2011}. In principle, astrometric observations favor longer period orbits; however, high precision needs to be maintained over the correspondingly longer time baselines. For shorter periods, the planets need to be massive enough to introduce a detectable astrometric wobble in the star and {\it Gaia} should begin to contribute here~\citep{Perryman2001}. Microlensing offers sensitivity to planets in wider orbits and has contributed to our statistical knowledge about occurrence rates of longer period planets~\citep[e.g., ][]{Gaudi2010,Cassan2012} and direct imaging of planets in wide orbits is also beginning to contribute important information~\citep{Oppenheimer2009}. 

Here, we announce \totalnum long-period transiting exoplanet candidates from the {\it Kepler} mission. These planet candidates mostly have 1-3 visible transits and typically have orbital periods between 100 and 2000 days, corresponding to orbital separations from their host stars of 1-3 AU. The candidate systems were identified by citizen scientists taking part in the Planet Hunters project\footnote{http://www.planethunters.org/}. We  obtain follow-up adaptive optics (AO) images (for \totalao host stars) and spectroscopic observations (for \eaespec host stars) in an effort to validate the planet candidates and characterize their host stars. We derive their orbital and stellar parameters by fitting transiting light curves and performing spectral classification.

The Planet Hunters project began in December 2010 as part of the Zooniverse\footnote{https://www.zooniverse.org} network of Citizen Science Projects. The project displays light curves from the {\it Kepler} mission to crowd-source the task of identifying transits ~\citep{Fischer2012}. This method is effective in finding potential exoplanets not flagged by the {\it Kepler} data reduction pipeline, since human classifiers can often spot patterns in data that would otherwise confuse computer algorithms. The detection efficiency of the volunteers is independent of the number of transits present in the light curve, i.e., they are as likely to identify a single transit as multiple transits in the same lightcurve, however the probability of identifying planets is higher if the transit is deeper.~\citet{Schwamb2012b} described the weighting scheme for transit classifications. ~\citet{Wang2013} and \citet{Schmitt2013} described the process of vetting planet candidates in detail as well as the available tools on the Planet Hunters website. 

The paper is organized as follows. In \S \ref{sec:methods}, we model transiting light curves of planet candidates and derive stellar and orbital properties of these candidate systems. In \S \ref{sec:followup}, we present adaptive optics (AO) imaging for \totalao systems and spectroscopic observations for \eaespec systems. In \S \ref{sec:individual}, {{we calculate planet confidence for each planet candidate }}and discuss notable candidate systems. Finally, we conclude in \S \ref{sec:summary} with a summary and discussions of future prospects. 


\section{Planet candidates and their host stars}
\label{sec:methods}

Planet Hunters identified \totalnum long-period planet candidates around \totalst stars. In this section, we describe the procedures with which we modeled these transit curves and estimated the stellar properties of their host stars. Since \totalsingle planet candidates exhibit only one visible transit, their orbital periods can not be well-determined. We provide a method of constraining the orbital period for a single-transit event based on transit duration and host star properties. 

\subsection{Modeling Light Curves}

We downloaded the {\it Kepler} light curves from the Mikulski Archive for Space Telescopes (MAST\footnote{http://archive.stsci.edu}) and detrended the quarterly segments using the autoKep software in the Transit Analysis Package~\citep[TAP,][]{Gazak2012}. The light curves were then modeled using TAP which adopts an analytic form for the model described by~\citet{Mandel2002}. The free parameters in the model include orbital period, eccentricity, argument of periastron, inclination, the ratio of semi-major axis and stellar radius $a/R_\ast$, the planet-star radius ratio $R_p/R_\ast$, mid transit time, linear and quadratic limb darkening parameters. We are particularly interested in $R_p/R_\ast$ and $a/R_\ast$. The former is used to determine the planet radius. The latter helps to estimate the orbital periods for planet candidates with only one visible transit. The following equation of constraining orbital period is derived based on Equation 18 and 19 from~\citet{Winn2010b}:
\begin{equation} 
\frac{P}{1\ \rm{yr}}=\left(\frac{T}{13\ \rm{hr}}\right)^3\cdot\left(\frac{\rho}{\rho_{\odot}}\right)\cdot(1-\rm{b}^2)^{-\frac{3}{2}},
\label{eq:per_est}
\end{equation}
where $P$ is period, $T$ is the transit duration, i.e., the interval between the halfway points of ingress and egress, $\rho$ is stellar density, $\rho_{\odot}$ is the solar density, and $b$ is the impact parameter. In a transit observation, the transit duration, $T$, is an observable that can be parametrized the follow way:
\begin{equation} 
T=\frac{1}{\pi}\cdot P\cdot\left(\frac{a}{R_\ast}\right)^{-1}\cdot\sqrt{(1-\rm{b}^2)}\cdot\frac{\sqrt{1-e^2}}{1+e\sin\omega},
\label{eq:t_dur}
\end{equation}
where $e$ is orbital eccentricity and $\omega$ is the argument of periastron. 

Most of planet candidates in this paper have orbital periods between 100 and 2000 days, and some of these are likely to be in eccentric orbits. Eccentricity affects the transit duration. For example, the transiting duration of a planet on an eccentric orbit can be longer than that for a circular orbit if viewed from the time of apastron. Unfortunately, it is very difficult to know whether long transit durations are caused by long orbital periods or high eccentricities, especially if the stellar radius is uncertain. However, since 80\% of known planets with orbital periods longer than 100 days have eccentricities lower than 0.3\footnote{http://exoplanets.org/}, we adopt a simplified prior assumption of zero eccentricity in our models.  This feeds into our estimates for orbital periods of those systems with only one transit, however the effect is not large. The main uncertainty for the planet period estimation comes from uncertainties in the stellar radius. For example, a typical 40\% stellar radius error translates to a $\sim$40\% $a/R_\ast$ error. Given that the observable $T$ stays the same, the 40\% stellar radius error leads to 40\% period estimation error according to Equation \ref{eq:t_dur}. In comparison, floating the eccentricity between 0 and 0.3 typically changes $P$ by 20\%. Therefore, the effect of eccentricity is smaller than the effect of stellar radius error on period estimation. Furthermore, setting eccentricity to zero reduces the number of free parameters by two, i.e., eccentricity and argument of periastron; this facilitates the convergence of the Markov chains in TAP analysis. This is especially useful when there are only 1-4 transits available to constrain the model. The posterior distribution of the MCMC analysis is used to contain the orbital period (\S \ref{sec:per}) for systems that only have a single transit.

\begin{figure}[htp]
\begin{center}
\includegraphics[angle=0, width= 0.9\textwidth]{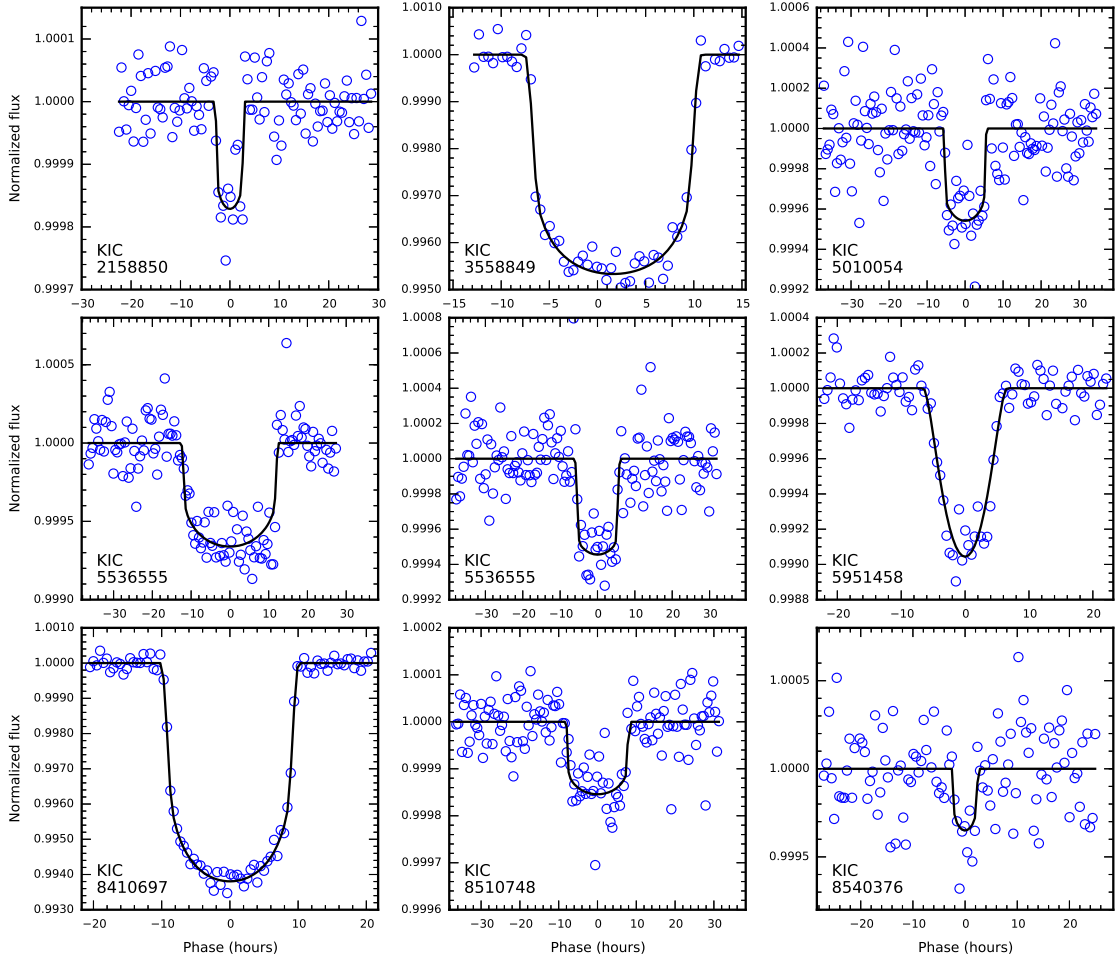} 
\caption{Transiting light curves for 1-transit planet candidates. Blue open circles are data points and black solid line is the best-fitting model. Orbital parameters can be found in Table \ref{tab:orbital_params_1}.
\label{fig:1transits1}}
\end{center}
\end{figure}

\begin{figure}[htp]
\begin{center}
\includegraphics[angle=0, width= 0.9\textwidth]{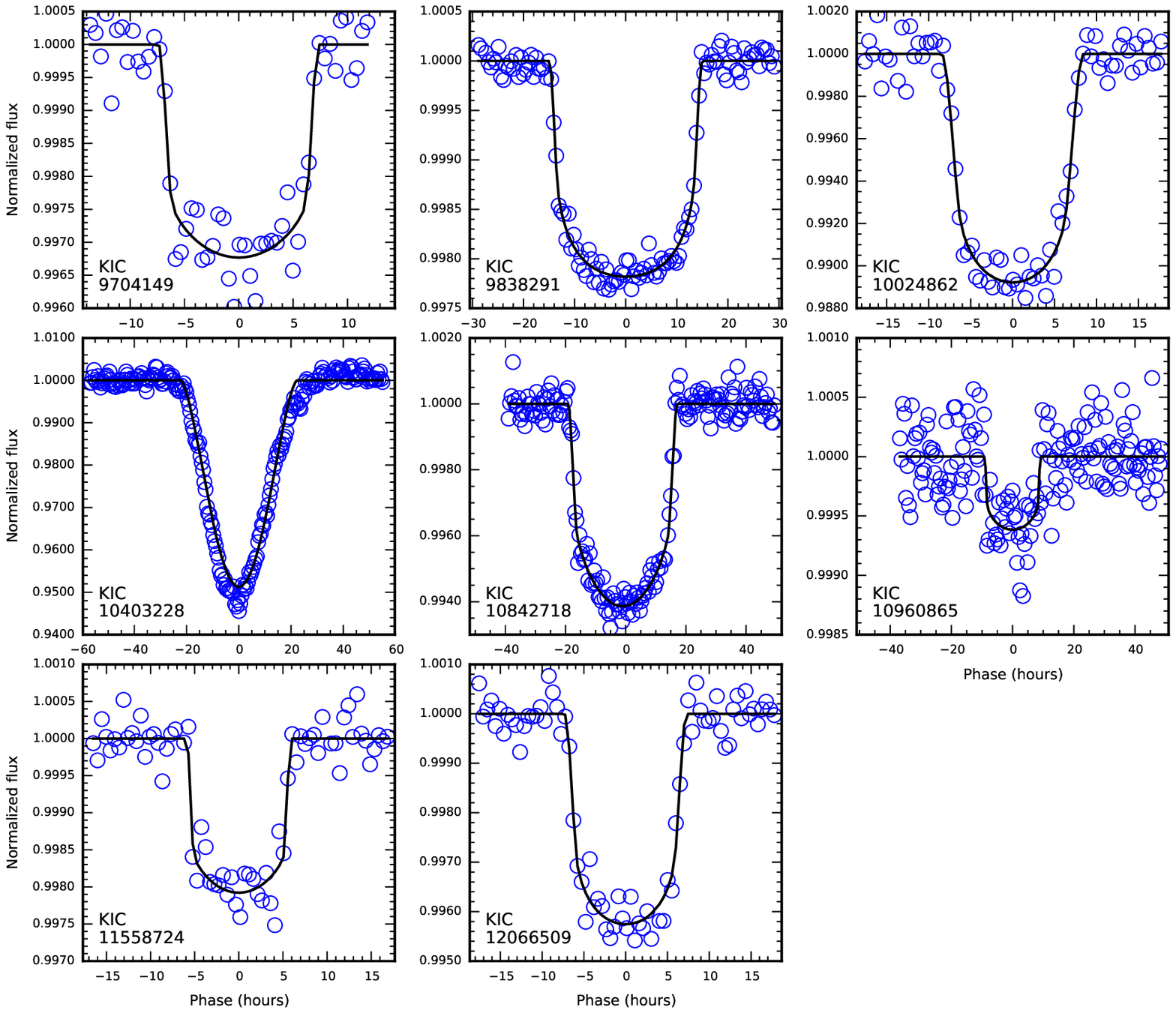} 
\caption{Transiting light curves for 1-transit planet candidates. Blue open circles are data points and black solid line is the best-fitting model. Orbital parameters can be found in Table \ref{tab:orbital_params_1}.
\label{fig:1transits2}}
\end{center}
\end{figure}

\begin{figure}[htp]
\begin{center}
\includegraphics[angle=0, width= 0.9\textwidth]{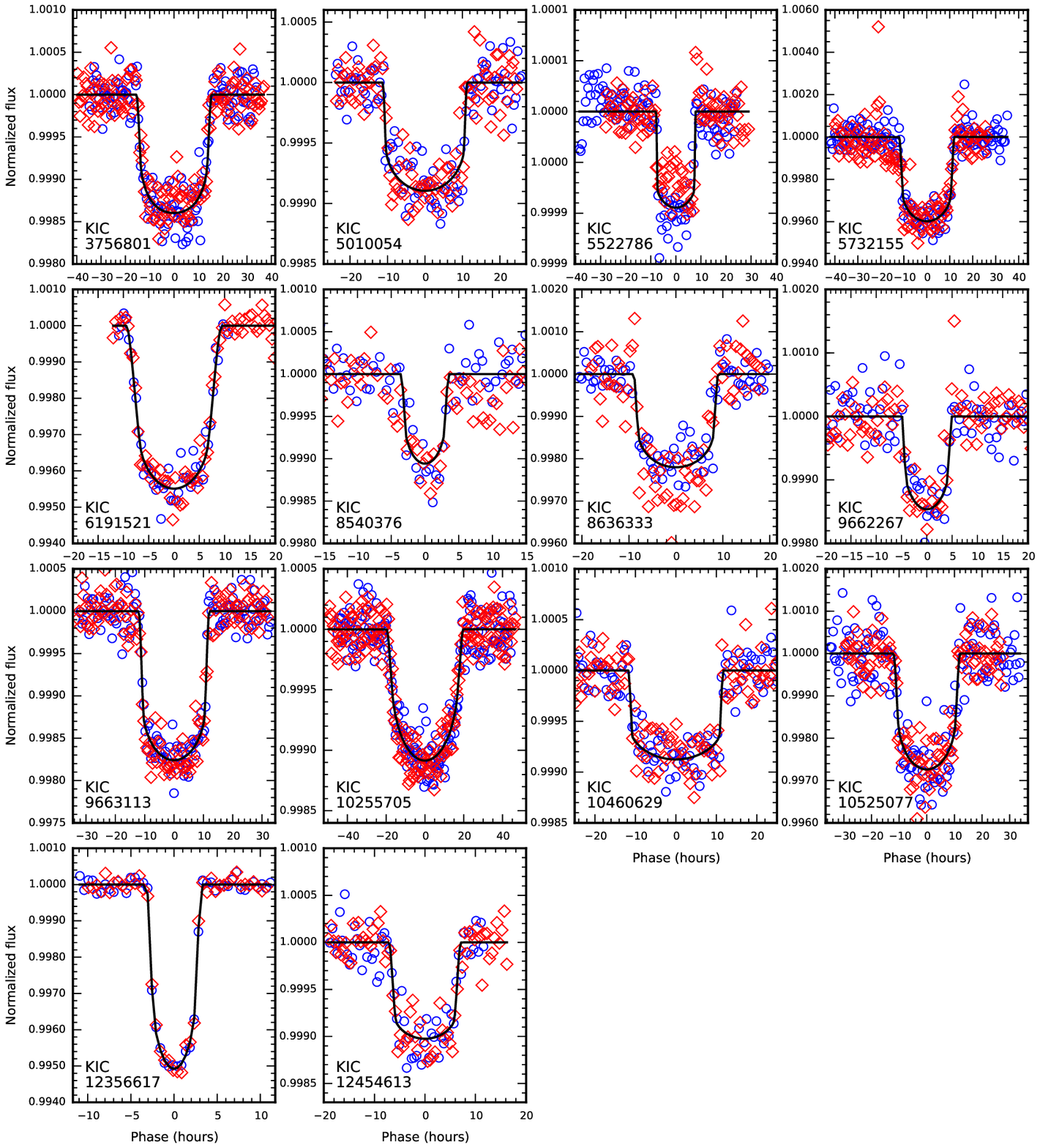} 
\caption{Transiting light curves for 2-transit planet candidates. Blue and red open circles are data points for odd- and even-numbered transits. Black solid line is the best-fitting model. Orbital parameters can be found in Table \ref{tab:orbital_params_2}.
\label{fig:2transits1}}
\end{center}
\end{figure}

\begin{figure}[htp]
\begin{center}
\includegraphics[angle=0, width= 0.9\textwidth]{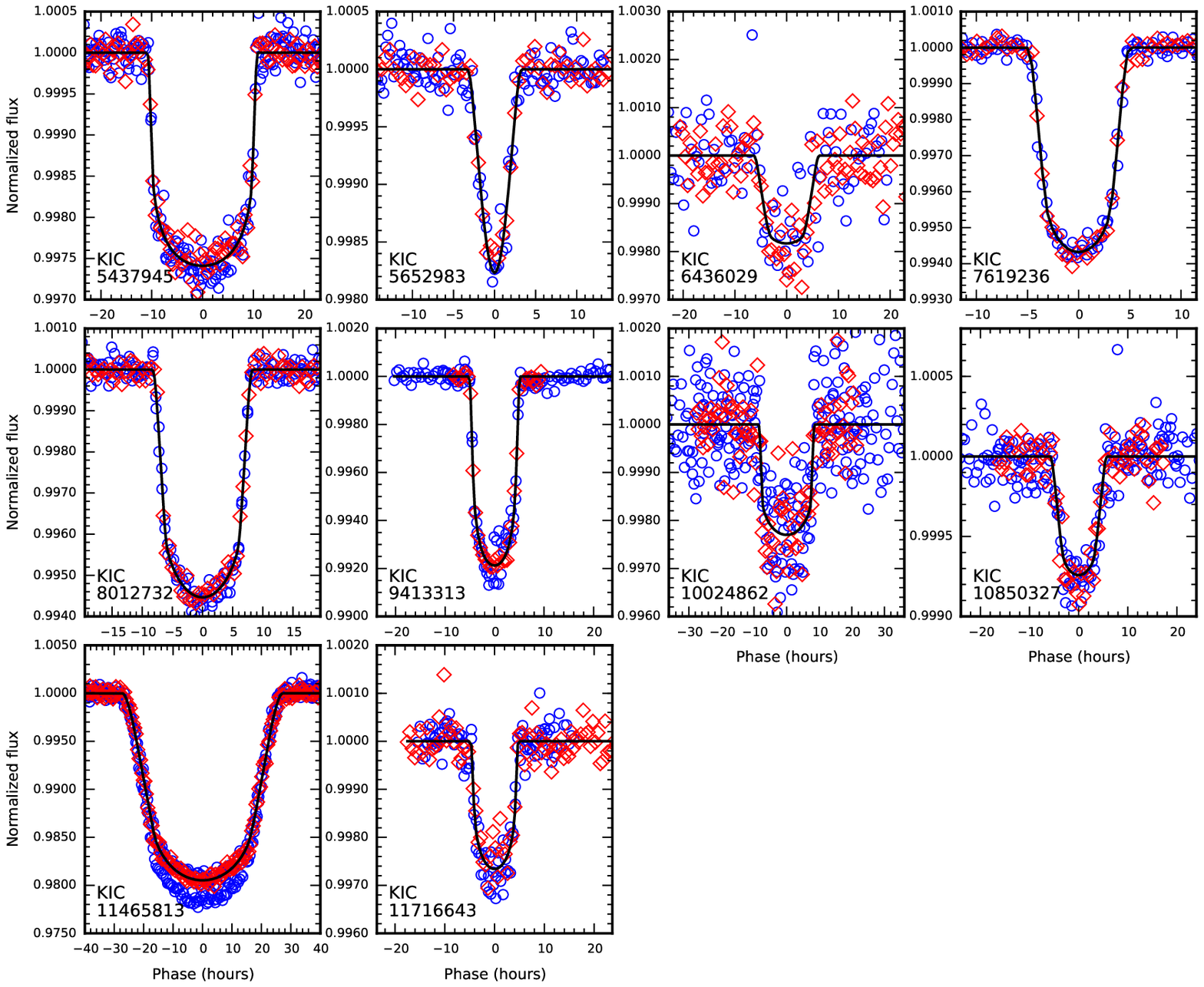} 
\caption{Transiting light curves for 3-transit planet candidates. Blue and red open circles are data points for odd- and even-numbered transits. Black solid line is the best-fitting model. Orbital parameters can be found in Table \ref{tab:orbital_params_3}. 
\label{fig:3transits}}
\end{center}
\end{figure}

We report results of light curve modeling for systems with only one observed transit (Table \ref{tab:orbital_params_1}, shown in Fig. \ref{fig:1transits1} and Fig. \ref{fig:1transits2}), two transits (Table \ref{tab:orbital_params_2}, shown in Fig. \ref{fig:2transits1}), and three transits (Table \ref{tab:orbital_params_3}, shown in Fig. \ref{fig:3transits}). 


\subsection{Stellar Mass and Radius}
\label{sec:star_mass_rad}

Characterizing host stars for planetary systems helps us to better understand the transiting planets. In particular, the planet radius can be calculated only if stellar radius is estimated. Stellar density is required for estimating the orbital periods for those planets that exhibit only one transit (see Equation \ref{eq:per_est}). We estimate stellar mass and radius in a similar way as~\citet{Wang2014a}: we infer these two stellar properties using the Yale-Yonsei Isochrone interpolator~\citep{Demarque2004}. The inputs for the interpolator are T$_{\rm{eff}}$, $\log g$, [Fe/H], $\alpha$ element abundance [$\alpha$/H] and stellar age. The first three parameters can be obtained by analyzing follow-up stellar spectra or from the NASA Exoplanet Archive\footnote{http://exoplanetarchive.ipac.caltech.edu} and the updated {\it Kepler} catalog for stellar properties~\citep{Huber2014}. We set [$\alpha$/H] to be the solar value, zero, and allow stellar age to vary between 0.08 and 15 Gyr. We ran a Monte Carlo simulation to consider measurement uncertainties of T$_{\rm{eff}}$, $\log g$, [Fe/H]. For stars with spectroscopic follow-up observations (\S \ref{sec:spec_fop}), the uncertainties are based on the MOOG spectroscopic analysis~\citep{Sneden1973}. For stars that are {\it Kepler} Objects of Interest (KOIs), the uncertainties are from the NASA Exoplanet Archive. 
We report  the 1$\sigma$ ranges for stellar masses and radii in Table \ref{tab:stellar_params} along with T$_{\rm{eff}}$, $\log g$, and [Fe/H].

\subsection{Orbital Period}
\label{sec:per}

Orbital periods are a fundamental parameter for exoplanets and are often used to understand the prospects for habitability. For systems with more than one visible transit, we determined the orbital period by calculating the time interval between transits. The uncertainty of the orbital period is calculated by propagating the measurement error of the mid transit time of each transit. For systems with only one visible transit, we use Equation \ref{eq:per_est} to estimate the orbital period $P$, as a function of the transit duration $T$, stellar density $\rho$, and the impact parameter b. $T$ and $b$ can be constrained by modeling the transiting light curve. For instance, $T$ can be measured directly from the transit observation, and $b$ can be inferred by fitting the light curve. On the other hand, $\rho$ can be constrained by stellar evolutionary model as described in \S \ref{sec:star_mass_rad}. Therefore, with knowledge of $T$, $\rho$, and $b$ from transit observation and stellar evolutionary model, we can constrain the orbital period for planet candidates with only a single transit. 

We start with a test TAP run to obtain the posterior distribution of the transit duration $T$ (Equation \ref{eq:t_dur}) and impact parameter b. The distribution of stellar density can be obtained from the process as described in \S \ref{sec:star_mass_rad}. We then start a Monte Carlo simulation to infer the distribution of orbital period. In the simulation, we sample from $T$, $b$ and $\rho$ distributions, which result in a distribution of orbital period. We report the mode and 1-$\sigma$ range of orbital period in Table \ref{tab:orbital_params_1}. 

\begin{figure}[htp]
\begin{center}
\includegraphics[angle=0, width= 0.9\textwidth]{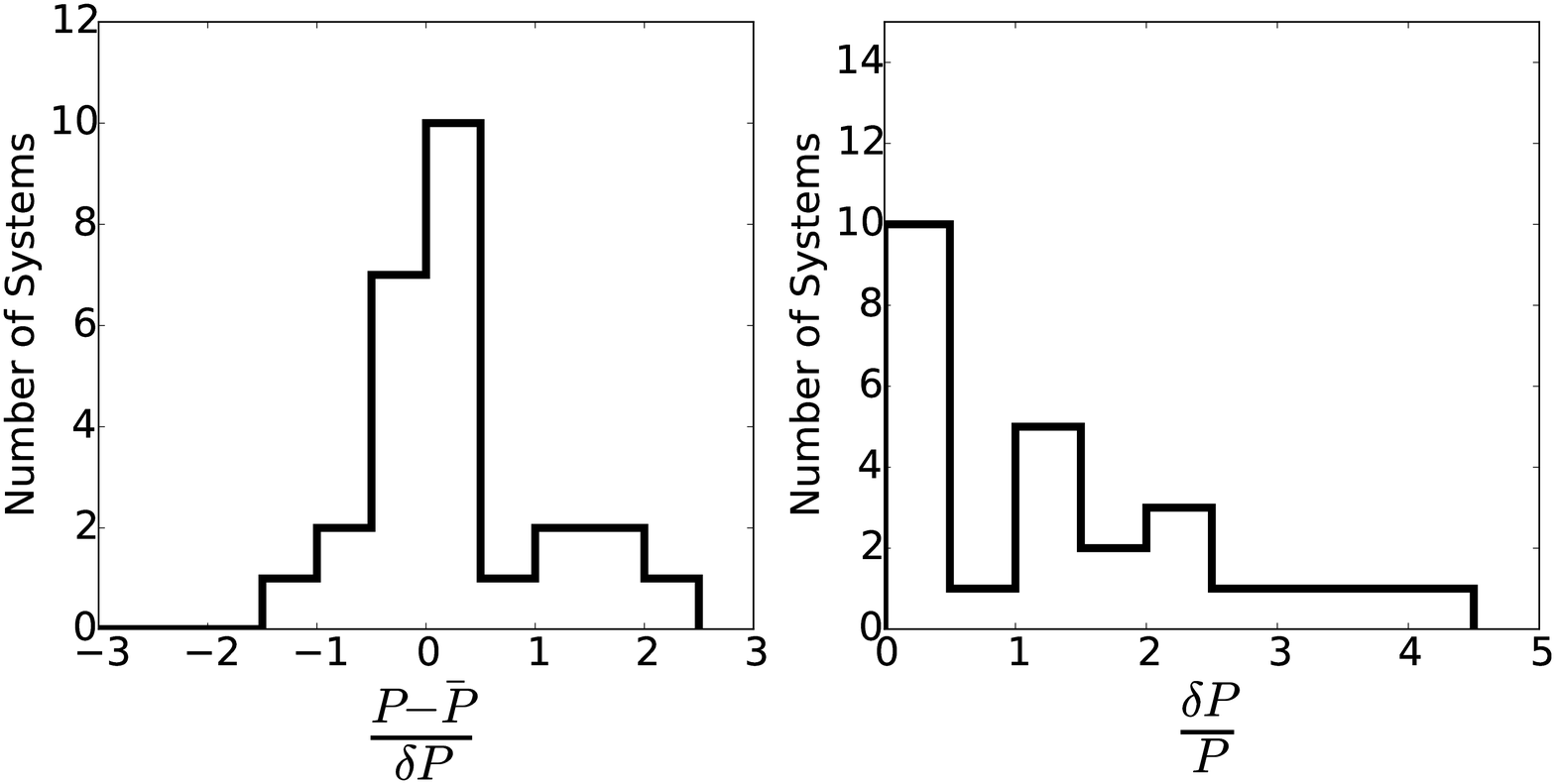} 
\caption{Left: distribution of the difference between the period estimated from individual transit ($\bar{P})$ and the period estimated from the time interval of consecutive transits ($P$) for 24 candidate planetary systems with 2-3 visible transits. The difference is normalized by measurement uncertainty of $\delta P$. Right: distribution of the fractional error $\delta P/P$. 
\label{fig:test}}
\end{center}
\end{figure}

We investigate the error of our period estimation using systems with known orbital periods. For the 24 planet candidates with 2-4 transits in this paper, we compare the period ($\bar{P}$) estimated from individual transit and the period ($P$) based on the interval between mid-transit, which is much more precise than $\bar{P}$. If $\bar{P}$ and $P$ are in agreement within 1-$\sigma$ error bars, then the method used for single-transit systems would seem to give a reasonable estimate and uncertainty for orbital period. The left panel of Fig. \ref{fig:test} shows the distribution of the difference between $\bar{P}$ and $P$ normalized by measurement uncertainty $\delta P$, which is calculated as half of the 1-$\sigma$ range from the Monte Carlo simulation. About 69\% of the comparisons are within 1-$\sigma$ range, which indicates that $\bar{P}$ and $P$ agree for the majority of cases and $\delta P$ is a reasonable estimation of measurement uncertainty. The right panel of Fig. \ref{fig:test} shows the fractional error ($\delta P$/$P$) distribution of the orbital periods estimated from individual transit. The median fractional error is 1.4 and the fractional error is smaller than 50\% for 34\% of cases, which suggests that the period estimated from an individual transit has a large uncertainty, i.e., hundreds of days. This is because of the weak dependence of transit duration on orbital period, i.e., $T\sim P^{1/3}$, a large range of $P$ would be consistent with the measured transit duration. As a result, orbital period uncertainty for systems with a single transit is much larger than systems with more than one visible transit. However, the estimation of orbital period provides a time window for follow-up observations.

\section{Follow-up Observations}
\label{sec:followup}

Follow-up observations include AO imaging and spectroscopy of host stars with planet candidates. AO imaging can identify additional stellar components in the system or in the foreground/background. These can be potential sources for flux contamination~\citep[e.g.,][]{Dressing2014} or false positives~\citep[e.g.,][]{Torres2011}. Spectroscopic follow-up observations are used to derive stellar properties that are more reliable than those derived with multi-band photometry. Furthermore, since follow-up observations exclude some scenarios for false positives, the likelihood of a planet candidate being a bona-fide planet can be increased and a planet candidate can be statistically validated~\citep[e.g.,][]{Barclay2013}. In this section, we describe our AO imaging and spectroscopic follow-up observations. In addition, we discuss sources from which we obtain archival data and information about these planet host stars. 

\subsection{Adaptive Optics Observations}

{{In total, AO images were taken for \totalao stars with planet candidates in this paper. We observed \keckao targets with the NIRC2 instrument~\citep{Wizinowich2000} at the Keck II telescope using the Natural Guiding Star mode. }}The observations were made on UT July 18th and August 18th in 2014, {{and August 27-28 in 2015}} with excellent/good seeing between 0.3$^{\prime\prime}$ to 0.9$^{\prime\prime}$. NIRC2 is a near infrared imager designed for the Keck AO system. We selected the narrow camera mode, {{which has a pixel scale of 9.952 mas pixel$^{-1}$~\citep{Yelda2010}}}. The field of view (FOV) is thus $\sim$10$^{\prime\prime}\times$10$^{\prime\prime}$ for a mosaic 1K $\times$1K detector. We started the observation in the $K_s$ band for each target. The exposure time was set such that the peak flux of the target is at most 10,000 ADU for each frame, which is within the linear range of the detector. We used a 3-point dither pattern with a throw of 2.5$^{\prime\prime}$. We avoided the lower left quadrant in the dither pattern because it has a much higher instrumental noise than other 3 quadrants on the detector. We continued observations of a target in $J$ and $H$ bands if any stellar companions were found. 

We observed 1 target with the PHARO instrument\citep{Brandl1997,Hayward2001} at the Palomar 200-inch telescope. The observation was made on UT July 13rd 2014 with seeing varying between 1.0$^{\prime\prime}$ and 2.5$^{\prime\prime}$. PHARO is behind the Palomar-3000 AO system, which provides an on-sky Strehl of up to 86\% in $K$ band~\citep{Burruss2014}. The pixel scale of PHARO is 25 mas pixel$^{-1}$. With a mosaic 1K $\times$1K detector, the FOV is 25$^{\prime\prime}\times$25$^{\prime\prime}$. We normally obtained the first image in the $K_s$ band with a 5-point dither pattern, which had a throw of 2.5$^{\prime\prime}$. The exposure time setting criterion is the same as the Keck observation: we ensured that the peak flux is at least 10,000 ADU for each frame. If a stellar companion was detected, we observed the target in $J$ and $H$ bands. 


We observed \roboao targets between UT 2014 Aug 23rd and 30th with the Robo-AO system installed on the 60-inch telescope at Palomar Observatory \citep{Baranec2013, Baranec2014}. Observations consisted of a sequence of rapid frame-transfer read-outs of an electron multiplying CCD camera with 0\farcs043 pixels at 8.6 frames per second with a total integration time of 90 s in a long-pass filter cutting on at 600nm. The images were reduced using the pipeline described in Law et al. (2014). In short, after dark subtraction and flat-fielding using daytime calibrations, the individual images were up-sampled, and then shifted and aligned by cross-correlating with a diffraction-limited PSF. The aligned images were then co-added together using the Drizzle algorithm \citep{Fruchter02} to form a single output frame. The final ``drizzled'' images have a finer pixel scale of 0\farcs02177/pixel.

The raw data from NIRC2 and PHARO were processed using standard techniques to replace bad pixels, flat-field, subtract thermal background, align and co-add frames. We calculated the 5-$\sigma$ detection limit as follows. We defined a series of concentric annuli centering on the star. For the concentric annuli, we calculated the median and the standard deviation of flux for pixels within these annuli. We used the value of five times the standard deviation above the median as the 5-$\sigma$ detection limit. We report the detection limit for each target in Table \ref{tab:ao_params}. Detected companions are reported in Table \ref{tab:ao_detection}. 

\subsection{Spectroscopic Observation}
\label{sec:spec_fop}

We obtained stellar spectra for \eaespec stars using the East Arm Echelle (EAE) spectrograph at the Palomar 200-inch telescope. The EAE spectrograph has a spectral resolution of $\sim$30,000 and covers the wavelength range between 3800 to 8600 \AA. The observations were made between UT Aug 15th and 21st 2014. The exposure time per frame is typically 30 minutes. We usually obtained 2-3 frames per star and bracketed each frame with Th-Ar lamp observations for wavelength calibration. Because these stars are faint with {\it Kepler} magnitudes mostly ranging from 13 to 15.5 mag, the signal to noise ratio (SNR) of their spectra is typically 20-50 per pixel at 5500 \AA. 



We used IDL to reduce the spectroscopic data to get wavelength calibrated, 1-d, normalized spectra. These spectra were then analyzed by the newest version of MOOG~\citep{Sneden1973} to derive stellar properties such as effective temperature (T$_{\rm{eff}}$), surface gravity ($\log g$) and metallicity [Fe/H]~\citep{Santos2004}. The iron line list used here was obtained from~\citet{Sousa2008} excluding all the blended lines in our spectra due to a limited spectral resolution. The measurement of the equivalent widths was done systematically by fitting a gaussian profile to the iron lines. The equivalent widths together with a grid of Kurucz Atlas 9 plane-parallel model atmospheres~\citep{Kurucz1993} were used by MOOG to calculate the ion abundances. The errors of the stellar parameters are estimated using the method described by~\citet{Gonzalez1998}. The targets with spectroscopic follow-up observations are indicated in Table \ref{tab:stellar_params}.

\subsection{Archival AO and Spectroscopic Data From CFOP}

For those targets for which we did not conduct follow-up observations, we searched the {\it Kepler} Community Follow-up Observation Program\footnote{https://cfop.ipac.caltech.edu} (CFOP) for archival AO and spectroscopic data. We found that only one target had AO images from CFOP. KIC-5857656 was observed at the Large Binocular Telescope on UT Oct 3rd 2014, but the image data was not available. A total of \cfopspec of targets had spectroscopic data based on CFOP, but only \cfopspect of them had uploaded stellar spectra. We used the spectroscopically-derived stellar properties for these stars in the subsequent analyses.

\subsection{Stars without AO and Spectroscopic Data}

For stars without AO and spectroscopic data, we obtained their stellar properties from the NASA Exoplanet Archive if they were identified as {\it Kepler} Objects of Interest (KOIs). If the stars are not KOIs, then we obtained their stellar properties from the update {\it Kepler} catalog for stellar properties~\citep{Huber2014}. 

\section{Planet Candidates and Notable Systems}
\label{sec:individual}

\begin{figure}[htp]
\begin{center}
\includegraphics[angle=0, width= 0.9\textwidth]{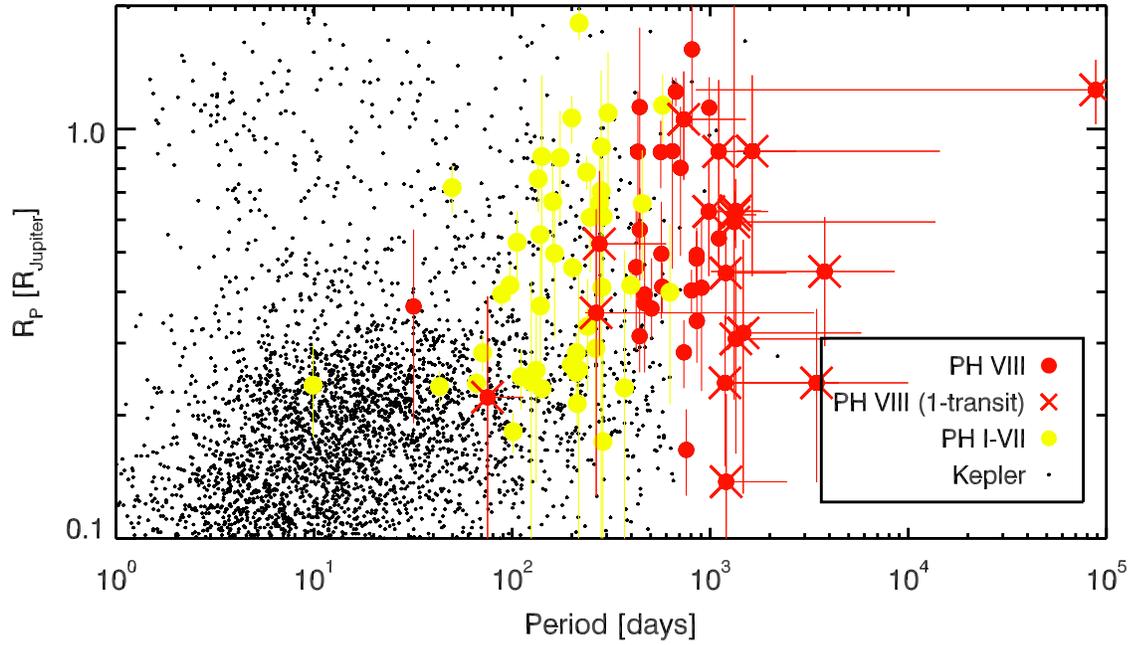} 
\caption{Scatter plot of planet radii vs. orbital periods for planet candidates discovered with {\it Kepler} data. Black dots are {\it Kepler} planet candidates. Red filled circles are planet candidates from this work that are identified by Planet Hunters. Planet candidates with a single transit are marked with red crosses. Yellow filled circles are planet candidates from previous Planet Hunters papers: PH I \citep{Fischer2012}, PH II: \citep{Schwamb2012}, PH III: \citep{Schwamb2013}, PH IV: \citep{Lintott2013}, PH V: \citep{Wang2013}, PH VI: \citep{Schmitt2013}, PH VII: \citep{Schmitt2014}. Long-period planet candidates are predominantly discovered by Planet Hunters. 
\label{fig:Per_Rp}}
\end{center}
\end{figure}

Fig. \ref{fig:Per_Rp} shows a scatter plot of planet radii and orbital periods for planet candidates found with the {\it Kepler} data. Most of the known KOIs (88\%) have orbital periods shorter than 100 days so the planet candidates discovered by the Planet Hunters help to extend the discovery space into the long period regime.  We emphasize that we have included in this paper planet candidates with one or two observed transits which would otherwise excluded by the {\it Kepler} pipeline. This approach enables the Planet Hunters project to be more sensitive to long-period planet candidates, allowing us to explore a larger parameter space. 

\subsection{Planet Confidence of Planet Candidates}
\label{sec:planet_confidence}

{{

The follow-up observations for these long-period candidates help to exclude false positive scenarios such as background or physically-associated eclipsing binaries. We use a method called planetary synthesis validation (PSV) to quantify the planet confidence for each planet candidate~\citep{Barclay2013}. PSV has been used to validate several planet candidates such as Kepler-69c~\citep{Barclay2013}, PH-2b~\citep{Wang2013}, and Kepler-102e~\citep{Wang2014a}. PSV makes use of transiting observations and follow-up observations to exclude improbable regions in parameter space for false positives. For parameter space that cannot be excluded by observations, PSV adopts a Bayesian approach to calculate the probability of possible false positives and gives an estimation of planet confidence between 0 and 1 with 1 being an absolute bona-fide planet. We adopt a planet confidence threshold of 0.997 (3-$\sigma$) for planet validation. The threshold is more conservative than previous works~\citep[e.g., ][]{Rowe2014}.

The inputs for the PSV code are planet radius, transit depth, pixel centroid offset between in and out of transit, pixel centroid offset significance (i.e., offset divided by measurement uncertainty), number of planet candidates, and the AO contrast curve of the host star in the absence of stellar companion detection. \citet{Wang2013} provided details in the procedures of deriving these inputs and the methodology for the PSV method. The output of the PSV code is the planet confidence, the ratio between planet prior and the sum of the planet prior and possible false positives. Table \ref{tab:planet_confidence} provides the results of PSV. There are 7 planet candidates that have planet confidences over 0.997. Their planet statuses are therefore validated. These planets include three planets with a single transit (KIC-3558849b or KOI-4307b, KIC-5951458b, and KIC-8540376c), 3 planets with double transits (KIC-8540376b, KIC-9663113b or KOI-179b, and KIC-10525077b or KOI-5800b), and 1 planet with 4 transits (KIC-5437945b or KOI-3791b). The notation for each planet (e.g., b, c, d) starts from the innermost planet candidate. 

Since this work contains single and double-transit planet candidates that are typically overlooked by the {\it Kepler} mission~\citep{Borucki2010,Borucki2011,Batalha2013,Burke2014}, it is informative to investigate the false positive rate for this population of planet candidates. Out of 24 candidate systems for which we have AO data, 6 have detected stellar companions (see Table \ref{tab:ao_detection}). Depending on which star hosts the transiting object, 4 systems may be false positives due to an underestimated radius (see further discussions in the following sections). These systems are KIC-8510748 (1-transit), KIC-8636333 (2-transit), KIC-11465813 (3-transit), and KIC-12356617 (2-transit). Two other systems have planet candidates whose radii remain in the planetary regime despite the flux dilution effect (KIC-5732155 and KIC-10255705). In addition, two candidate systems have planet confidences lower than 0.85 (KIC-10024862 and KIC-11716643). If considering all candidate systems with detected stellar companions and candidate systems with planet confidences lower than 0.85 as false positives, an aggressive estimation of the false positive rate for single or double-transit planet candidates is 33\%. If considering only the 4 candidates systems that may be false positives due to flux dilution, a conservative estimation of the false positive rate is 17\%. 
}}

\subsection{Single-Transit Systems}
\label{sec:single}



\textbf{3558849}
This star is listed as KOI-4307 and has one planet candidate with period of 160.8 days, but KOI-4307.01 does not match with the single transit event. Therefore this is an additional planet candidate in the same system. {{The additional single-transit planet candidate KIC-3558849b is validated with a planet confidence of 0.997. }}


\textbf{5010054}
This target is not in the threshold crossing event (TCE) or KOI tables. Three visible transits are attributed to two planet candidates. The first two at BKJD 356 and 1260~\citep{Schmitt2013} are from the same object (they are included in the following Double-Transit Systems section). The third transit at BKJD 1500 is different in both transit depth and duration, so it is modeled here as a single transit from a second planet in the system.


\textbf{5536555}
There are two single-transit events for this target (BKJD 370 and 492). We flag the one at BKJD 370 as a cosmic-ray-induced event. It is caused by Sudden Pixel Sensitivity Dropout~\citep[SPSD,][]{Christiansen2013,Kipping2015}. After a cosmic ray impact, a pixel can lose its sensitivity for hours, which mimics a single-transit event. A cosmic ray hitting event is marked as a SAP\_QUALITY 128 event when cosmic ray hits pixels within photometric aperture and marked as a  SAP\_QUALITY 8192 event when cosmic ray hits adjacent pixels of a photometric aperture. The single-transit event at BKJD 370 coincides with with a SAP\_QUALITY 128 event, so we caution that it may be an artifact. However, the single-transit event at BKJD 492 is still a viable candidate. Single-transit events that are caused by SPSD are also found for other {\it Kepler} stars. We list here the SPSDs found by Planet Hunters and the associated BKJDs: KIC-9207021 (BKJD 679), KIC-9388752 (BKJD 508), and KIC-10978025 (BKJD 686). 

\textbf{5951458}
{{This planet candidate KIC-5951458b is validated with a planet confidence of 0.998. }}

\textbf{8540376}
There are only two quarters of data for this target (Q16 and Q17). However, there are three planet candidates in this system. One starts at BKJD 1499.0 and has an orbital period of 10.7 days. One has only two observed transits with an orbital period of 31.8 days. The two-transit system will be discussed in the following section (\S \ref{sec:double}). There is a single-transit event (BKJD 1516.9), which appears to be independent of the previous two planet candidates. This single-transit event would be observed again soon because its orbital period has a 1-$\sigma$ upper limit of 114.1 days. {{The planet candidate exhibiting single transit (KIC-8540376c) is validated with a planet confidence of 0.999. }}

%

\textbf{9704149}
There is a second possible transit at BKJD 1117, but only ingress is recorded here and the rest of the transit is lost due to a data gap. If the second transit is due to the same object, then the orbital period is 697.3 days, which is at odds with the estimated period at 1199.3 days.

\textbf{10024862}
In addition to the single transit event, there is also a second object with three visible transits (P = 567.0 days, see \S \ref{sec:triple}). The triple-transit system was also reported in~\citet{Wang2013}, but there were only two visible transits at that time.

\textbf{10403228}
This is a transit event from a planet around a M dwarf. Despite the deep transit ($\sim$5\%), the radius of the transiting object is within planetary range (R$_P=9.7$ R$_{\oplus}$). However, the transit is v-shaped, suggesting a grazing transit and the true nature of the transiting object is uncertain. For the M star, we adopt stellar mass and radius from~\citet{Huber2014} which uses the Dartmouth stellar evolutionary model~\citep{Dotter2008}.  

\textbf{10842718}
The orbital period distribution given the constraints from transit duration and stellar density (\S \ref{sec:per}) has two peaks. One is at $\sim$1630 days, the other one is at $\sim$10,000 days. The bimodal distribution suggests that the orbital period of this transiting object could be much longer than reported in Table \ref{tab:orbital_params_1}, however the probability for a transiting planet with a period of $\sim$10,000 days is vanishingly low, giving stronger weight to the shorter period peak.  


\subsection{Double-Transit Systems}
\label{sec:double}

\textbf{3756801}
This object is first mentioned in~\citet{Batalha2013} and designated as KOI-1206. Surprisingly, it appears that only one transit was detected by \citet{Batalha2013}. It does not appear in the {\it Kepler} TCE table because a third transit was not observed. 

\textbf{5732155}
A stellar companion has been detected in $K_S$ band that is 4.94 magnitudes fainter. The separation of the stars is 1$^{\prime\prime}$ (Table \ref{tab:ao_detection}). The flux contamination does not significantly change the transit depth and thus does not affect planet radius estimation. The stellar companion is so faint that even a total eclipsing binary would not yield the observed transit depth.  

\textbf{6191521}
This target is listed as KOI-847 and has one planet candidate with orbital period of 80.9 days. Here, we report a second, longer-period planet candidate that was not previously detected in the system. 

\textbf{8540376}
There are only two quarters of data for this target (Q16 and Q17), but there are three planet candidates in this system.  The longer period single-transit event has been discussed in \S \ref{sec:individual}. The double-transit event starts at BKJD 1520.3 and has a period of 31.8 days. The shortest period planet (10.7 days) has transits that begin at BKJD 1499.0. {{The double-transit planet candidate KIC-8540376b is validated with a planet confidence of 0.999. }}

\textbf{8636333}
This target is listed as KOI-3349 and has two planet candidates. One is KOI-3349.01 with period of 82.2 days; the other one was reported in~\citet{Wang2013} with period of 804.7 days. Here, we report the follow-up observations for this star: a fainter stellar companion has been detected in $H$ and $K_S$ bands (Table \ref{tab:ao_detection}) with differential magnitudes of 1.58 and 1.71 in these filters respectively. We estimate their {\it Kepler} band magnitudes to be different by $\sim$ 3 mag. Based on Fig. 11 in~\citet{Horch2014}, a correction for the radius of the planet that accounts for flux from the stellar companion would increase the planet radius by a small amount, $\sim$3\%. However, if the two candidates are transiting the fainter secondary star, then their radii would increase by a factor of $\sim$3. In this case, although the radii for both candidates would remain in planetary range, the longer-period candidate would be at the planetary radius threshold.


\textbf{9663113}
This target is listed as KOI-179 and has two planet candidates. One is KOI-179.01 with period of 20.7 days. KOI-179.02 was reported in~\citet{Wang2013} with period of 572.4 days with two visible transits. The expected third transit at BKJD 1451 is missing, but the expected position is in a data gap. {{The double-transit planet candidate KIC-9663113b is validated with a planet confidence of 0.999. }}

\textbf{10255705}
This target was reported in~\citet{Schmitt2013}. Follow-up AO observation shows that there is a nearby stellar companion (Table \ref{tab:ao_detection}). The companion is $\sim$2 mag fainter in {\it Kepler} band. If the planet candidate orbits the primary star, then the planet radius adjustment due to flux contamination is small. If the planet candidate orbits around the newly detected stellar companion, then the planet radius is revised upward by a factor of $\sim$2~\citep{Horch2014}, but the adjusted radius is still within planetary range. 

\textbf{10460629}
This target is listed as KOI-1168 and has one planet candidate that matches with the double-transit event. There are two deep v-shaped dips in the lightcurve at BKJD 608.3 and 1133.3, likely indicating an eclipsing binary within the planet orbit. These v-shaped transits are so deep (about 13\%) that they could easily be followed up from the ground. If the planet interpretation is correct for the other two transit events, then this could be an circumbinary planet candidate. However, this system is likely to be a blending case in which two stars are in the same photometric aperture. This system is discussed further in \S \ref{sec:circumbinary}.

\textbf{10525077}
This target is listed as KOI-5800 and has one planet candidate with period of 11.0 days. The second planet candidate was reported in~\citet{Wang2013} with period of 854.1 days. There are two transits at BKJD 355.2 and 1189.3. In between these two transits, there is a data gap at 762.3, preventing us from determining whether the orbital period is 854.1 days or half of the value, i.e, 427.05 days. {{This planet candidate KIC-10525077b is validated with a planet confidence of 0.998. }}

\textbf{12356617}
This target is listed as KOI-375 and has one planet candidate that matches with the double-transit event. Follow-up AO observation shows that there is one faint stellar companion at 3.12$^{\prime\prime}$ separation. If the transit occurs for the primary star, the radius adjustment due to flux contamination is negligible. If the transit occurs for the secondary star, then this is a false positive. 

\subsection{Triple or Quadruple Transit Systems}
\label{sec:triple}

%
%

\textbf{5437945}
This target is listed as KOI-3791and has two planet candidates in 2:1 resonance. KOI-3791.01 was reported in~\citet{Wang2013} and~\citet{Huang2013}. The fourth transit appears at BKJD 1461.8. {{The longer-period planet candidate KIC-5437945b  is validated with a planet confidence of 0.999. }}

\textbf{5652983}
This target is listed as KOI-371 and has one planet candidate that matches with the triple-transit event. The radius of the transiting object is too large to be a planet, and thus the triple-transit event is a false positive, which is supported by the notes from CFOP that large RV variation has  been observed. 


\textbf{6436029}
This target is listed as KOI-2828 and has two planet candidates. KOI-2828.02 with period of 505.5 days matches the triple-transit event. KOI-2828.02 was reported in~\citet{Schmitt2013}, but there were only two visible transits.

\textbf{7619236}
This target is listed as KOI-5205 and has one planet candidate that matches with the triple-transit event. It exhibits significant transit timing variations (TTVs). The time interval between the first two transits is different by $\sim$27 hours from the time interval between the second and the third transit.  

\textbf{8012732}
This object was reported in~\citet{Wang2013}. It exhibits significant TTVs. The time interval between the first two transits is different by $\sim$20 hours from the time interval between the second and the third transit.  

\textbf{9413313}
This object was reported in~\citet{Wang2013}. It exhibits significant TTVs. The time interval between the first two transits is different by $\sim$30 hours from the time interval between the second and the third transit.  

\textbf{10024862}
This object was reported in~\citet{Wang2013}, but only two transit were observed then. The third transit is observed at BKJD 1493.8.  It exhibits significant TTVs. The time interval between the first two transits is different by $\sim$41 hours from the time interval between the second and the third transit.   

\textbf{10850327}
This target is listed as KOI-5833 and has one planet candidate that matches with the triple-transit event. The object  was reported in~\citet{Wang2013}, but there were only two transit observed then.

\textbf{11465813}
This target is listed as KOI-771 and has one planet candidate that matches with the triple-transit event. The transit depth is varying. This target also has a single transit at BKJD 1123.5. A stellar companion has been detected (Table \ref{tab:ao_detection}). From the colors of the companion, we estimate the differential magnitude to be 0.7 mag. The radius of the object would be revised upward by 23\% or 150\% depending on whether the object orbits the primary or the secondary star. In either case, it is likely that this object is a false positive. 

\textbf{11716643}
This target is listed as KOI-5929 and was reported in~\citet{Wang2013}, but there were only two transits observed then. It exhibits TTVs. The time interval between the first two transits is different by $\sim$2.7 hours from the time interval between the second and the third transit.  

\subsection{Notable False Positives}
\label{sec:false_positives}

In addition to the systems with single-transit events flagged as SPSDs in \S \ref{sec:single}, we list other transiting systems that are likely to be false positives.

\textbf{1717722}
This target is listed as KOI-3145 with two known planet candidates. Neither candidate matches the single transit event at BKJD 1439. This single transit is likely spurious, as pixel centroid offset between in- and out-of-transit are seen for this transit. 

\textbf{3644071}
This target is listed as KOI-1192 and has one false positive (02) and one candidate (01). The epoch for candidate KOI-1192.01 matches with the epoch of the single transit event in this paper. According to notes on CFOP, the KOI-1192 event is ``due to video crosstalk from an adjacent CCD readout channel of the image of a very bright, highly saturated star". This effect causes the varying transit depth and duration. The explanation is further supported by the apparent pixel offset between in- and out-of-transit for both KOI-1192.01 and KOI-1192.02. So KOI-1192.01 is also likely to be a false positive. 

\textbf{9214713}
This target is listed as KOI-422 and has one planet candidate that matches with the double-transit event found by Planet Hunters. {{Significant pixel centroid offset is found between in- and out-of-transit although follow-up AO and spectroscopic observations show no sign of nearby stellar companions. }}

\textbf{10207400}
This target is currently not in either the Kepler KOI or TCE tables. There is a pixel centroid offset between in- and out-of-transit.

\section{Summary and Discussion}
\label{sec:summary}

\subsection{Summary}
We report \totalnum long-period planet candidates around \totalst {\it Kepler} stars. These planet candidates are identified by the Planet Hunters based on the archival {\it Kepler} data from Q0 to Q17. We conduct AO imaging observations to search for stellar companions and exclude false positive scenarios such as eclipsing binary blending. In total, we obtain AO images for \totalao stars. We detect stellar companions around 6 stars within 4$^{\prime\prime}$, KIC-5732155, KIC-8510748, KIC-8636333 (KOI-3349), KIC-10255705, KIC-11465813 (KOI-771), and KIC-12356617 (KOI-375). The properties of these stellar companions are given in Table \ref{tab:ao_detection}. For those stars with non-detections, we provide AO sensitivity limits at different angular separations (Table \ref{tab:ao_params}). We obtain high-resolution spectra for a total of \eaespec stars. We use the stellar spectra to infer stellar properties such as stellar mass and radius which are used for orbital period estimation for single-transit events. The stellar properties of planet host stars are given in Table \ref{tab:stellar_params}. We model the transiting light curves with TAP to obtain their orbital parameters. Table \ref{tab:orbital_params_1}, Table \ref{tab:orbital_params_2} and Table \ref{tab:orbital_params_3} give the results of light curve modeling for single-transit, double-transit, and triple/quardruple-transit systems, respectively. Based on transiting and follow-up observations, we calculate the planet confidence for each planet candidate. Seven planet candidates have planet confidence above 0.997 and are thus validated. These planets include 3 planets with a single transit (KIC-3558849b or KOI-4307b, KIC-5951458b, and KIC-8540376c), 3 planets with double transits (KIC-8540376b, KIC-9663113b or KOI-179b, and KIC-10525077b or KOI-5800b), and 1 planet with 4 transits (KIC-5437945b or KOI-3791b). We estimate the false positive rate to be 17\%-33\% for 1-transit and 2-transit events. 

\subsection{KIC-10460629: An Interesting Case}
\label{sec:circumbinary}

{{At first glance, KIC-10460629 might be an extreme circumbinary planetary system if confirmed with a $P=856.7$ days planet and a $P=525$ days eclipsing binary star. The ratio of semi-major axis of the transiting planet to the eclipsing secondary star is $\sim$1.4. The tight orbital configuration makes the system dynamically unstable. According to Equation 3 in~\citet{HW99}, the minimum semi-major axis ratio for a stable orbit around a binary star is 2.3 for a binary with $e=0$ and $\mu=0.5$, where $\mu$ is the mass ratio of the primary to the secondary star estimated from the transit depth (13\%). Therefore, KIC-10460629 should be dynamical unstable. Furthermore, the minimum semi-major axis ratio increases with increasing eccentricity, which makes the systems even more unstable for eccentric orbits based on the criterion from ~\citet{HW99}. }}

{{A more likely explanation for the observed two sets of transits is a blending case, in which two stars are within the photometric aperture and each has one set of transits. There are two possibilities in the blending case. For the first case (referred to as Case A), the deep transit takes place around the brighter star and the shallower transit takes place around the fainter star. In this case, the fainter star needs to brighter than 7.8 differential magnitude in the {\it Kepler} band, otherwise it does not produce the observed 784 ppm transit even with a total ecplise. Our AO observations are not deep enough to rule out this scenario. For second possible blending case (referred to as Case B), the deep transit takes place around the fainter star and the shallower transit takes place around the brighter star. In this case, the fainter star needs to be brighter than 2.2 differential magnitude in the {\it Kepler} band in order to produce the observed 13\% transit depth. This possibility is ruled out by our AO observations for angular separations larger than 0$^{\prime\prime}$.1. The blending has to happen within 0$^{\prime\prime}$.1 angular separation for Case B. There might be a Case C, in which the two shallower transits are not caused by the same object. However, there is no evidence that this is the case given the similarity of the two transits (see Fig. \ref{fig:2transits1}).  
}}

{{Follow-up observations are necessary to determine the nature of this transiting system. For Case A, a deeper AO observation is required to confirm or rule out the fainter star. For Case B, a high-resolution spectroscopy of the target would reveal the fainter source since it is at least 13\% as bright as the brighter star. Long time-baseline RV observations can also differentiate the two cases. For Case A, a clear stellar RV signal should be observed. In case B, the precision of RV measurements may not be adequate to map out the orbit of the transiting planet candidate with a Neptune-size given the low-mass and the faintness of the host star (K$_P=14.0$). Ground-based transiting follow-up observations can certainly catch the transit of the secondary star at 13\% depth. The next transit of the secondary star will on UT June 1st 2016. The transit depth of the planet candidate is $\sim$800 ppm, which may be detected by ground-based telescopes. The next shallower transit will be on UT August 30 2016. This may confirm or rule out Case C.   }}

\subsection{Evidence of Additional Planets in Systems with Long-Period Transiting Planets}

{{TTVs indicate the likely presence of additional components in the same system that dynamically interacting with transiting planet candidates. For the \totaltriple systems with 3-4 visible transits for which we can measure TTVs, 50\% (5 out of 10) exhibit TTVs ranging from $\sim$2 to 40 hours. Four systems have synodic TTV larger than 20 hours, making them the ``queens" of transit variations as opposed to the 12-hr ``king" system KOI-142~\citep{Nesvorny2013}. Excluding two likely false positives, KIC-5652983 (large RV variation) and KIC-11465813 (blending), the fraction of systems exhibiting TTVs goes up to 68\%. All such systems host giant planet candidates with radii ranging from 4.2 to 12.6 $\rm{R}_{\oplus}$. Based on Equation 10 in~\citet{Deck2015}, or Fig. 6 in~\citet{Nesvorny2014}, an order of magnitude estimation for the mass of the perturber is a few Jupiter masses, assuming an orbital separation corresponding to a period ratio $\lesssim 2$, and low eccentricity orbits. In general, to maintain the same amplitude synodic TTV, the mass of the perturber would need to increase for wider separations, and decrease for smaller. }}

This result suggests that most long-period transiting planets have at least one additional companion in the same system. This finding is consistent with the result in ~\citet{Fischer2001} that almost half (5 out of 12) of gas giant planet host stars exhibit coherent RV variations that are consistent with additional companions. This finding is further supported by a more recent study of companions to systems with hot Jupiters~\citep{Knutson2014,Ngo2015}, in which the stellar and planetary companion rate of hot Jupiter systems is estimated to be $\sim$50\%. While we emphasize the different planet populations between previous studies (short-period planets) and systems reported in this paper (long-period planets), the companion rate for stars with gas giant planets is high regardless of the orbital period of a planet. 

~\citet{Dawson2013} found that giant planets orbiting metal-rich stars show signatures of planet-planet interactions, suggesting that multi-planet systems tend to favorably reside in metal-rich star systems. We check the metallicities of the five systems exhibiting TTVs. The median metallicity is $0.07\pm0.18$. In comparison, the median metallicity for the entire sample is $-0.06\pm0.38$ dex. While there is a hint that the TTV sample is more metal rich, the large error bars and the small sample prevent us from further studying the metallicity distribution of systems exhibiting TTVs. However, studying the metallicity of planet host stars remains a viable tool and future follow-up observations would allow us to use the tool to test planet formation theory.

\subsection{The Occurrence Rate of Long-Period Planets}
The presence of long-period planets may affect the evolution of multi-planet systems by dynamical interaction~\citep[e.g.,][]{Rasio1996,Dong2014}. The dynamical effects result in observable effects such as spin-orbit misalignment which provides constraints on planet migration and evolution~\citep[e.g.,][]{Winn2010}. Therefore, measuring the occurrence rate of long-period planets is essential in determine their role in planet evolution.~\citet{Cumming2008} estimated that the occurrent rate is 5-6\% per period decade for long-period gas giant planets. ~\citet{Knutson2013} estimated that $51\%\pm10\%$ of hot-Jupiter host stars have an additional gas giant planet in the same system. However, these studies are sensitive to planets with mass higher than $\sim$0.3 Jupiter mass. The {\it Kepler} mission provides a large sample of small planets (likely to be low-mass planets), which can be used to infer the occurrence rate for small, long-period planets. However, such analysis is limited to periods up to $\sim$500 days~\citep{Dong2013,Petigura2013,Rowe2015}. The upper limit is due to the 3-transit detection criterion for {\it Kepler} planet candidates. With the long-period planet candidates in this paper, we will be able to probe the occurrence rate of planets between 1 and 3 AU. To accomplish this goal, a proper assessment of planet recovery rate of the Planet Hunters is required. The framework has already been provided by~\citet{Schwamb2012b} and this issue will be addressed in a future paper.  Estimating the occurrence rate of Neptune to Jupiter-sized planets between 1 and 3 AU will be an important contribution of Planet Hunters to the exoplanet community. 

\subsection{K2 and TESS}

The current K2 mission and future TESS (Transiting Exoplanet Survey Satellite) missions have much shorter continuous time coverage than the {\it Kepler} mission. Each field of the K2 mission receives $\sim$75 days continuous observation~\citep{Howell2014}. For the TESS mission, the satellite stays in the same field for 27.4 days~\citep{Ricker2015}. Despite longer time coverage for a portion of its field, the majority of sky coverage of TESS will receive only 27.4 days observation. Given the scanning strategy of these two missions, there will be many single-transit events. Estimating the orbital periods for these events is crucial if some the targets with a single transit have significant scientific value, e.g., planets in the habitable zone. More generally, estimating orbital period helps to predict the next transit and facilitates follow-up observations, especially for those searching for the next transit~\citep{Yee2008}. Once more than one transits are observed, more follow-up observations can be scheduled such as those aiming to study transiting planets in details, e.g., CHEOPS (CHaracterising ExOPlanet Satellite) and JWST (James Webb Space Telescope). 

\noindent{\it Acknowledgements} 
The authors would like to thank the anonymous referee whose comments and suggestions greatly improve the paper. We are grateful to telescope operators and supporting astronomers at the Palomar Observatory and the Keck Observatory. Some of the data presented herein were obtained at the W.M. Keck Observatory, which is operated as a scientific partnership among the California Institute of Technology, the University of California and the National Aeronautics and Space Administration. The Observatory was made possible by the generous financial support of the W.M. Keck Foundation. The research is made possible by the data from the {\it Kepler} Community Follow-up Observing Program (CFOP). The authors acknowledge all the CFOP users who uploaded the AO and RV data used in the paper. We thank Katherine M. Deck for insightful comments on TTV systems and the dynamical stability of KIC-10460629. This research has made use of the NASA Exoplanet Archive, which is operated by the California Institute of Technology, under contract with the National Aeronautics and Space Administration under the Exoplanet Exploration Program. JW, DF and TB acknowledge the support from NASA under Grant No. NNX12AC01G and NNX15AF02G.

The Robo-AO system was developed by collaborating partner institutions, the California Institute of Technology and the Inter-University Centre for Astronomy and Astrophysics, and with the support of the National Science Foundation under Grant No. AST-0906060, AST-0960343 and AST-1207891, the Mt. Cuba Astronomical Foundation and by a gift from Samuel Oschin.C.B. acknowledges support from the Alfred P. Sloan Foundation. KS gratefully acknowledges support from Swiss National Science Foundation Grant PP00P2\_138979/1

Facilities: 
PO:1.5m (Robo-AO)
PO: 5m (PHARO)
KO: 10m (NIRC2)

\bibliography{mybib_AP_LPPaper}



\newpage

\clearpage

\begin{landscape}
\begin{deluxetable}{lclllccllcll}
\tabletypesize{\tiny}
 \setlength{\tabcolsep}{0.02in} 
\tablewidth{0pt}
\tablecaption{Orbital Parameters (1 visible transit)\label{tab:orbital_params_1}}
\tablehead{
\colhead{\textbf{KIC}} &
\colhead{\textbf{KOI}} &
\colhead{\textbf{P Mode}} &
\colhead{\textbf{P Range}} &
\colhead{\textbf{$a/\rm{R}_\ast$}} &
\colhead{\textbf{Inclination}} &
\colhead{\textbf{$\rm{R}_P/\rm{R}_\ast$}} &
\colhead{\textbf{$\rm{R}_P$}} &
\colhead{\textbf{Epoch}} &
\colhead{\textbf{$\mu_1$}} &
\colhead{\textbf{$\mu_2$}} &
\colhead{\textbf{Comments}} \\
\colhead{\textbf{}} &
\colhead{\textbf{}}	&
\colhead{\textbf{(days)}}	&
\colhead{\textbf{(days)}} &
\colhead{\textbf{ }}	&
\colhead{\textbf{(deg)}}	&
\colhead{\textbf{}}	&
\colhead{\textbf{($\rm{R}_\oplus$)}}	&
\colhead{\textbf{(BKJD)}} &
\colhead{\textbf{}} &
\colhead{\textbf{}} &
\colhead{\textbf{See \S \ref{sec:single} for details}}
}

\startdata

2158850$^\dagger$ & & 1203.8 & [1179.3..2441.6] & $1037.9^{+225.7}_{-274.6}$ & $89.956^{+0.058}_{-0.110}$ & $0.013^{+0.002}_{-0.001}$ & $1.6^{+1.0}_{-0.8}$ & $411.791^{+0.008}_{-0.008}$ & $0.520^{+0.330}_{-0.350}$ & $-0.070^{+0.450}_{-0.430}$ & \\
3558849$^\dagger$ & 04307 & 1322.3 & [1311.1..1708.4] & $576.7^{+21.5}_{-50.2}$ & $89.973^{+0.023}_{-0.031}$ & $0.063^{+0.002}_{-0.002}$ & $6.9^{+1.0}_{-0.9}$ & $279.920^{+0.440}_{-0.300}$ & $0.350^{+0.300}_{-0.230}$ & $0.220^{+0.370}_{-0.460}$ & Multi, Validated\\
5010054$^\dagger$ &  & 1348.2 & [1311.2..3913.9] & $825.1^{+134.8}_{-264.2}$ & $89.963^{+0.140}_{-0.250}$ & $0.021^{+0.002}_{-0.002}$ & $3.4^{+1.8}_{-1.6}$ & $1500.902^{+0.008}_{-0.009}$ & $0.400^{+0.380}_{-0.280}$ & $-0.040^{+0.420}_{-0.440}$ & Multi\\
5536555$^\dagger$ & & 3444.7 & [1220.7..9987.4] & $908.4^{+119.4}_{-191.1}$ & $89.965^{+0.024}_{-0.038}$ & $0.024^{+0.003}_{-0.003}$ & $2.7^{+1.4}_{-1.1}$ & $370.260^{+0.033}_{-0.038}$ & $0.500^{+0.350}_{-0.330}$ & $0.070^{+0.430}_{-0.420}$  &Cosmic artifact\\
5536555$^\dagger$ & & 1188.4 & [1098.6..4450.9] & $431.8^{+71.1}_{-97.8}$ & $89.887^{+0.068}_{-0.130}$ & $0.024^{+0.002}_{-0.001}$ & $2.7^{+1.2}_{-1.0}$ & $492.410^{+0.009}_{-0.008}$ & $0.510^{+0.340}_{-0.340}$ & $-0.140^{+0.410}_{-0.410}$  &\\
5951458$^\dagger$ & & 1320.1 & [1167.6..13721.9] & $278.1^{+109.1}_{-66.0}$ & $89.799^{+0.090}_{-0.120}$ & $0.040^{+0.089}_{-0.008}$ & $6.6^{+26.3}_{-4.2}$ & $423.463^{+0.010}_{-0.013}$ & $0.520^{+0.330}_{-0.350}$ & $0.000^{+0.430}_{-0.420}$  &Validated\\
8410697$^\dagger$ & & 1104.3 & [1048.9..2717.8] & $446.0^{+7.7}_{-17.1}$ & $89.976^{+0.024}_{-0.031}$ & $0.072^{+0.001}_{-0.001}$ & $9.8^{+4.9}_{-4.7}$ & $542.122^{+0.001}_{-0.001}$ & $0.410^{+0.130}_{-0.130}$ & $0.210^{+0.240}_{-0.230}$  &\\
8510748$^{\dagger,\dagger\dagger}$ & & 1468.3 & [1416.0..5788.4] & $569.5^{+145.4}_{-203.5}$ & $89.938^{+0.044}_{-0.073}$ & $0.012^{+0.001}_{-0.001}$ & $3.6^{+2.4}_{-2.1}$ & $1536.548^{+0.013}_{-0.015}$ & $0.500^{+0.300}_{-0.300}$ & $0.000^{+0.450}_{-0.450}$  &Binary\\
8540376$^\dagger$ & & 75.2 & [74.1..114.1] & $103.9^{+14.1}_{-14.1}$ & $89.701^{+0.160}_{-0.160}$ & $0.018^{+0.004}_{-0.005}$ & $2.4^{+1.9}_{-1.4}$ & $1516.911^{+0.020}_{-0.020}$ & $0.510^{+0.340}_{-0.340}$ & $0.000^{+0.440}_{-0.420}$  &Multi, Validated, Q16 and Q17 data only\\
9704149$^\dagger$ & & 1199.3 & [1171.3..2423.2] & $600.9^{+71.2}_{-121.8}$ & $89.955^{+0.059}_{-0.076}$ & $0.054^{+0.003}_{-0.003}$ & $5.0^{+1.4}_{-1.3}$ & $419.722^{+0.007}_{-0.007}$ & $0.490^{+0.330}_{-0.320}$ & $-0.080^{+0.450}_{-0.440}$  &Possible incomplete second transit\\
9838291$^\dagger$ & & 3783.8 & [1008.5..8546.1] & $930.4^{+72.1}_{-97.5}$ & $89.974^{+0.069}_{-0.063}$ & $0.043^{+0.001}_{-0.001}$ & $5.1^{+1.8}_{-1.8}$ & $582.559^{+0.003}_{-0.004}$ & $0.280^{+0.250}_{-0.180}$ & $0.430^{+0.280}_{-0.370}$  &\\
10024862$^\dagger$ & & 735.7 & [713.0..1512.8] & $324.1^{+41.2}_{-36.2}$ & $89.905^{+0.030}_{-0.022}$ & $0.098^{+0.004}_{-0.004}$ & $11.8^{+3.7}_{-3.4}$ & $878.561^{+0.004}_{-0.004}$ & $0.370^{+0.310}_{-0.250}$ & $0.280^{+0.370}_{-0.500}$ &Multi\\
10403228 & & 88418.1 & [846.5..103733.3] & $13877.4^{+400.0}_{-408.4}$ & $89.996^{+0.011}_{-0.009}$ & $0.269^{+0.022}_{-0.024}$ & $9.7^{+2.4}_{-2.2}$ & $744.843^{+0.013}_{-0.013}$ & $0.550^{+0.310}_{-0.370}$ & $0.050^{+0.420}_{-0.410}$ & V-shape\\
10842718 & & 1629.2 & [1364.7..14432.2] & $347.5^{+19.8}_{-23.9}$ & $89.938^{+0.010}_{-0.008}$ & $0.071^{+0.002}_{-0.002}$ & $9.9^{+5.4}_{-5.0}$ & $226.300^{+1.100}_{-0.520}$ & $0.700^{+0.180}_{-0.240}$ & $-0.150^{+0.400}_{-0.300}$ & Bimodal in inferred period\\
10960865 & & 265.8 & [233.7..3335.9] & $99.7^{+13.7}_{-28.6}$ & $89.703^{+0.240}_{-0.530}$ & $0.024^{+0.003}_{-0.003}$ & $3.9^{+3.0}_{-2.5}$ & $1507.959^{+0.007}_{-0.006}$ & $0.510^{+0.330}_{-0.340}$ & $-0.010^{+0.450}_{-0.450}$ & \\
11558724 & & 276.1 & [267.0..599.3] & $181.1^{+10.1}_{-25.8}$ & $89.897^{+0.021}_{-0.032}$ & $0.043^{+0.002}_{-0.002}$ & $5.9^{+2.9}_{-2.7}$ & $915.196^{+0.003}_{-0.003}$ & $0.470^{+0.330}_{-0.310}$ & $-0.130^{+0.460}_{-0.430}$ & \\
12066509 & & 984.6 & [959.0..1961.7] & $460.8^{+89.4}_{-72.3}$ & $89.925^{+0.050}_{-0.036}$ & $0.062^{+0.003}_{-0.003}$ & $7.1^{+2.3}_{-2.2}$ & $632.090^{+0.004}_{-0.004}$ & $0.360^{+0.330}_{-0.250}$ & $0.240^{+0.390}_{-0.500}$ & \\

\enddata

\tablecomments{$^\dagger$: Targets with AO follow-up observations. $^{\dagger\dagger}$: Targets with detected stellar companions as reported in Table \ref{tab:ao_detection}. The AO detection limits are given in Table \ref{tab:ao_params}. }

\end{deluxetable}

\clearpage
 \end{landscape}
\newpage

\clearpage

\begin{landscape}
\begin{deluxetable}{lcllccllcll}
\tabletypesize{\tiny}
 \setlength{\tabcolsep}{0.02in} 
\tablewidth{0pt}
\tablecaption{Orbital Parameters (2 visible transits)\label{tab:orbital_params_2}}
\tablehead{
\colhead{\textbf{KIC}} &
\colhead{\textbf{KOI}} &
\colhead{\textbf{P}} &
\colhead{\textbf{$a/\rm{R}_\ast$}} &
\colhead{\textbf{Inclination}} &
\colhead{\textbf{$\rm{R}_P/\rm{R}_\ast$}} &
\colhead{\textbf{$\rm{R}_P$}} &
\colhead{\textbf{Epoch}} &
\colhead{\textbf{$\mu_1$}} &
\colhead{\textbf{$\mu_2$}} &
\colhead{\textbf{Comments}} \\
\colhead{\textbf{}} &
\colhead{\textbf{}}	&
\colhead{\textbf{(days)}}	&
\colhead{\textbf{ }}	&
\colhead{\textbf{(deg)}}	&
\colhead{\textbf{}}	&
\colhead{\textbf{($\rm{R}_\oplus$)}}	&
\colhead{\textbf{(BKJD)}} &
\colhead{\textbf{}} &
\colhead{\textbf{}} &
\colhead{\textbf{See \S \ref{sec:double} for details}}
}

\startdata

3756801 & 01206 & $422.91360^{+0.01608}_{-0.01603}$ & $92.2^{+21.0}_{-27.0}$ & $89.620^{+0.280}_{-0.360}$ & $0.036^{+0.003}_{-0.002}$ & $5.1^{+2.2}_{-1.9}$ & $448.494^{+0.008}_{-0.008}$ & $0.260^{+0.310}_{-0.180}$ & $0.410^{+0.340}_{-0.500}$ & \\
5010054$^{\dagger}$ &  & $904.20180^{+0.01339}_{-0.01212}$ & $291.9^{+26.0}_{-62.0}$ & $89.918^{+0.057}_{-0.093}$ & $0.028^{+0.001}_{-0.001}$ & $4.6^{+2.2}_{-2.0}$ & $356.412^{+0.009}_{-0.008}$ & $0.460^{+0.330}_{-0.310}$ & $0.050^{+0.440}_{-0.450}$ & \\
5522786$^{\dagger}$ &  & $757.09520^{+0.01176}_{-0.01211}$ & $330.3^{+45.0}_{-77.0}$ & $89.913^{+0.062}_{-0.083}$ & $0.009^{+0.001}_{-0.001}$ & $1.9^{+0.4}_{-0.3}$ & $282.995^{+0.009}_{-0.008}$ & $0.320^{+0.360}_{-0.230}$ & $-0.060^{+0.360}_{-0.400}$ &  \\
5732155$^{\dagger,\dagger\dagger}$ &  & $644.21470^{+0.01424}_{-0.01598}$ & $204.3^{+15.0}_{-31.0}$ & $89.894^{+0.073}_{-0.100}$ & $0.059^{+0.002}_{-0.002}$ & $9.9^{+5.2}_{-4.7}$ & $536.702^{+0.006}_{-0.005}$ & $0.410^{+0.320}_{-0.270}$ & $0.040^{+0.440}_{-0.460}$  & Binary\\
6191521 & 00847 & $1106.24040^{+0.00922}_{-0.00954}$ & $326.6^{+30.0}_{-26.0}$ & $89.862^{+0.020}_{-0.020}$ & $0.068^{+0.002}_{-0.002}$ & $6.0^{+0.8}_{-0.6}$ & $382.949^{+0.007}_{-0.008}$ & $0.480^{+0.340}_{-0.320}$ & $0.150^{+0.390}_{-0.400}$ & Multi \\
8540376 & & $31.80990^{+0.00919}_{-0.00933}$ & $34.7^{+3.7}_{-6.9}$ & $89.300^{+0.490}_{-0.720}$ & $0.030^{+0.002}_{-0.002}$ & $4.1^{+2.2}_{-1.9}$ & $1520.292^{+0.006}_{-0.006}$ & $0.570^{+0.300}_{-0.350}$ & $0.030^{+0.440}_{-0.420}$ & Multi, Validated, Q16 and Q17 data only\\
8636333$^{\dagger\dagger}$ & 03349 & $804.71420^{+0.01301}_{-0.01500}$ & $343.8^{+21.0}_{-52.0}$ & $89.946^{+0.038}_{-0.062}$ & $0.044^{+0.002}_{-0.002}$ & $4.5^{+0.5}_{-0.5}$ & $271.889^{+0.009}_{-0.012}$ & $0.420^{+0.340}_{-0.280}$ & $0.080^{+0.430}_{-0.470}$ & Multi, Binary\\
9662267$^{\dagger}$ &  & $466.19580^{+0.00850}_{-0.00863}$ & $357.1^{+37.0}_{-82.0}$ & $89.931^{+0.049}_{-0.081}$ & $0.035^{+0.002}_{-0.002}$ & $4.5^{+1.7}_{-1.6}$ & $481.883^{+0.006}_{-0.006}$ & $0.590^{+0.280}_{-0.350}$ & $-0.060^{+0.460}_{-0.440}$  &\\
9663113 & 00179 & $572.38470^{+0.00583}_{-0.00567}$ & $153.5^{+23.0}_{-15.0}$ & $89.768^{+0.095}_{-0.062}$ & $0.041^{+0.001}_{-0.001}$ & $4.6^{+0.6}_{-0.7}$ & $306.506^{+0.004}_{-0.004}$ & $0.450^{+0.330}_{-0.270}$ & $0.040^{+0.390}_{-0.420}$ & Multi, Validated\\
10255705$^{\dagger,\dagger\dagger}$ &  & $707.78500^{+0.01844}_{-0.01769}$ & $92.1^{+27.0}_{-11.0}$ & $89.510^{+0.210}_{-0.110}$ & $0.034^{+0.002}_{-0.003}$ & $8.9^{+3.6}_{-3.5}$ & $545.741^{+0.014}_{-0.013}$ & $0.620^{+0.250}_{-0.350}$ & $0.250^{+0.350}_{-0.280}$ & Binary\\
10460629 & 01168 & $856.67100^{+0.01133}_{-0.01039}$ & $275.4^{+15.0}_{-40.0}$ & $89.932^{+0.048}_{-0.075}$ & $0.028^{+0.001}_{-0.001}$ & $3.8^{+0.8}_{-0.8}$ & $228.451^{+0.008}_{-0.006}$ & $0.420^{+0.320}_{-0.280}$ & $-0.070^{+0.420}_{-0.420}$ & EB, Unstable, Likely blending (\S \ref{sec:circumbinary}) \\
10525077 & 05800 & $854.08300^{+0.01628}_{-0.01697}$ & $239.3^{+46.0}_{-52.0}$ & $89.861^{+0.096}_{-0.098}$ & $0.050^{+0.003}_{-0.003}$ & $5.5^{+0.9}_{-0.8}$ & $335.236^{+0.012}_{-0.012}$ & $0.500^{+0.310}_{-0.310}$ & $0.140^{+0.410}_{-0.430}$ & Multi, Validated, Uncertain period (P = 427 or 854 days)\\
10525077 & 05800 & $427.04150^{+0.01487}_{-0.01628}$ & $130.9^{+14.0}_{-30.0}$ & $89.800^{+0.140}_{-0.220}$ & $0.049^{+0.003}_{-0.002}$ & $5.4^{+0.9}_{-0.8}$ & $335.238^{+0.011}_{-0.012}$ & $0.500^{+0.310}_{-0.310}$ & $0.160^{+0.410}_{-0.440}$  & Multi, Validated, Uncertain period (P = 427 or 854 days)\\
12356617$^{\dagger\dagger}$ & 00375 & $988.88111^{+0.00137}_{-0.00146}$ & $1059.5^{+29.0}_{-53.0}$ & $89.966^{+0.003}_{-0.004}$ & $0.069^{+0.001}_{-0.001}$ & $12.5^{+2.4}_{-2.3}$ & $239.224^{+0.001}_{-0.001}$ & $0.650^{+0.230}_{-0.320}$ & $-0.050^{+0.460}_{-0.330}$ & Binary\\
12454613$^{\dagger}$ &  & $736.37700^{+0.01531}_{-0.01346}$ & $257.0^{+140.0}_{-50.0}$ & $89.820^{+0.120}_{-0.064}$ & $0.033^{+0.002}_{-0.002}$ & $3.2^{+0.6}_{-0.6}$ & $490.271^{+0.014}_{-0.012}$ & $0.460^{+0.360}_{-0.310}$ & $-0.030^{+0.450}_{-0.430}$ & \\

\enddata

\tablecomments{All targets have AO imaging observations. Targets with follow-up spectroscopic observations are marked with a $^{\dagger}$. $^{\dagger\dagger}$: Targets with detected stellar companions as reported in Table \ref{tab:ao_detection}. The AO detection limits are given in Table \ref{tab:ao_params}. }

\end{deluxetable}

\clearpage
 \end{landscape}

\newpage

\clearpage

\begin{landscape}
\begin{deluxetable}{lcllccllcll}
\tabletypesize{\tiny}
 \setlength{\tabcolsep}{0.02in} 
\tablewidth{0pt}
\tablecaption{Orbital Parameters (3-4 visible transits)\label{tab:orbital_params_3}}
\tablehead{
\colhead{\textbf{KIC}} &
\colhead{\textbf{KOI}} &
\colhead{\textbf{P}} &
\colhead{\textbf{$a/\rm{R}_\ast$}} &
\colhead{\textbf{Inclination}} &
\colhead{\textbf{$\rm{R}_P/\rm{R}_\ast$}} &
\colhead{\textbf{$\rm{R}_P$}} &
\colhead{\textbf{Epoch}} &
\colhead{\textbf{$\mu_1$}} &
\colhead{\textbf{$\mu_2$}} &
\colhead{\textbf{Comments}} \\
\colhead{\textbf{}} &
\colhead{\textbf{}}	&
\colhead{\textbf{(days)}}	&
\colhead{\textbf{ }}	&
\colhead{\textbf{(deg)}}	&
\colhead{\textbf{}}	&
\colhead{\textbf{($\rm{R}_\oplus$)}}	&
\colhead{\textbf{(BKJD)}} &
\colhead{\textbf{}} &
\colhead{\textbf{}} &
\colhead{\textbf{See \S \ref{sec:triple} for details}}
}

\startdata

5437945 & 03791 & $440.78130^{+0.00563}_{-0.00577}$ & $158.9^{+5.1}_{-12.0}$ & $89.904^{+0.066}_{-0.086}$ & $0.047^{+0.001}_{-0.001}$ & $6.4^{+1.6}_{-1.6}$ & $139.355^{+0.003}_{-0.003}$ & $0.320^{+0.180}_{-0.160}$ & $0.290^{+0.270}_{-0.290}$ & Multi, Validated \\
5652983 & 00371 & $498.38960^{+0.01166}_{-0.01131}$ & $215.8^{+29.0}_{-33.0}$ & $89.721^{+0.049}_{-0.072}$ & $0.111^{+0.061}_{-0.057}$ & $35.9^{+24.7}_{-20.0}$ & $244.083^{+0.008}_{-0.008}$ & $0.550^{+0.310}_{-0.360}$ & $0.000^{+0.420}_{-0.430}$ & Large RV variation, likely a false positive \\
6436029 & 02828 & $505.45900^{+0.04500}_{-0.04102}$ & $155.5^{+32.0}_{-39.0}$ & $89.661^{+0.072}_{-0.150}$ & $0.047^{+0.012}_{-0.005}$ & $4.1^{+1.3}_{-0.7}$ & $458.092^{+0.035}_{-0.031}$ & $0.510^{+0.340}_{-0.360}$ & $0.000^{+0.430}_{-0.440}$ & Multi \\
7619236$^\dagger$ & 00682 & $562.70945^{+0.00411}_{-0.00399}$ & $311.9^{+16.0}_{-14.0}$ & $89.851^{+0.012}_{-0.011}$ & $0.077^{+0.002}_{-0.002}$ & $9.9^{+1.9}_{-1.8}$ & $185.997^{+0.002}_{-0.002}$ & $0.410^{+0.370}_{-0.280}$ & $0.230^{+0.360}_{-0.440}$ & TTV  \\
8012732$^\dagger$ & & $431.46810^{+0.00358}_{-0.00365}$ & $160.2^{+5.4}_{-4.6}$ & $89.741^{+0.018}_{-0.015}$ & $0.074^{+0.001}_{-0.002}$ & $9.8^{+4.1}_{-3.9}$ & $391.807^{+0.002}_{-0.002}$ & $0.560^{+0.290}_{-0.320}$ & $0.000^{+0.430}_{-0.360}$ & TTV \\
9413313$^\dagger$ & & $440.39840^{+0.00275}_{-0.00282}$ & $352.1^{+7.2}_{-15.0}$ & $89.966^{+0.023}_{-0.028}$ & $0.080^{+0.001}_{-0.001}$ & $12.6^{+7.2}_{-6.9}$ & $485.608^{+0.002}_{-0.002}$ & $0.380^{+0.130}_{-0.130}$ & $0.450^{+0.200}_{-0.250}$ & TTV \\
10024862$^\dagger$ & & $567.04450^{+0.02557}_{-0.02936}$ & $230.9^{+37.0}_{-81.0}$ & $89.868^{+0.095}_{-0.190}$ & $0.046^{+0.003}_{-0.003}$ & $5.5^{+2.0}_{-1.6}$ & $359.666^{+0.017}_{-0.021}$ & $0.410^{+0.370}_{-0.280}$ & $0.070^{+0.430}_{-0.480}$ & TTV \\
10850327 & 05833 & $440.16700^{+0.01738}_{-0.01671}$ & $124.9^{+36.0}_{-21.0}$ & $89.570^{+0.120}_{-0.100}$ & $0.032^{+0.003}_{-0.003}$ & $3.5^{+0.7}_{-0.6}$ & $470.358^{+0.011}_{-0.011}$ & $0.570^{+0.310}_{-0.370}$ & $0.120^{+0.390}_{-0.360}$ &  \\
11465813$^{\dagger\dagger}$ & 00771 & $670.65020^{+0.01018}_{-0.01018}$ & $85.2^{+1.1}_{-1.1}$ & $89.535^{+0.013}_{-0.012}$ & $0.136^{+0.002}_{-0.002}$ & $13.8^{+1.1}_{-1.1}$ & $209.041^{+0.004}_{-0.004}$ & $0.420^{+0.260}_{-0.240}$ & $0.340^{+0.360}_{-0.380}$ & Multi, Binary, Varying depth, Likely a false positive \\
11716643$^\dagger$ & 05929 & $466.00010^{+0.00799}_{-0.00775}$ & $380.5^{+24.0}_{-61.0}$ & $89.947^{+0.037}_{-0.059}$ & $0.047^{+0.002}_{-0.002}$ & $4.2^{+0.5}_{-0.4}$ & $434.999^{+0.005}_{-0.005}$ & $0.490^{+0.280}_{-0.290}$ & $0.230^{+0.370}_{-0.420}$ & TTV \\

\enddata

\tablecomments{All targets have AO imaging observations. Targets marked with a $^\dagger$ are systems displaying TTVs. $^{\dagger\dagger}$: Targets with detected stellar companions as reported in Table \ref{tab:ao_detection}. The AO detection limits are given in Table \ref{tab:ao_params}. }

\end{deluxetable}

\clearpage
 \end{landscape}
\newpage

\clearpage

\begin{deluxetable}{lcccccccccc}
\tabletypesize{\tiny}
 \setlength{\tabcolsep}{0.01in} 
\tablewidth{0pt}
\tablecaption{Stellar Parameters\label{tab:stellar_params}}
\tablehead{
\colhead{\textbf{KIC}} &
\colhead{\textbf{KOI}} &
\colhead{\textbf{$\alpha$}} &
\colhead{\textbf{$\delta$}} &
\colhead{\textbf{$Kp$}} &
\colhead{\textbf{$T_{\rm eff}$}} &
\colhead{\textbf{$\log g$}} &
\colhead{\textbf{[Fe/H]}} &
\colhead{\textbf{M$_{\ast}$}} &
\colhead{\textbf{R$_{\ast}$}} &
\colhead{\textbf{Orbital Solutions}}	\\
\colhead{\textbf{}} &
\colhead{\textbf{}} &
\colhead{\textbf{(h m s)}} &
\colhead{\textbf{(d m s)}} &
\colhead{\textbf{(mag)}}	&
\colhead{\textbf{(K)}}	&
\colhead{\textbf{(cgs)}}	&
\colhead{\textbf{(dex)}}	&
\colhead{\textbf{(M$_{\odot}$)}} &
\colhead{\textbf{(R$_{\odot}$)}}  &
\colhead{\textbf{in Table}} 
}

\startdata

2158850 &   & 19 24 37.875 & +37 30 55.69 &   10.9 & $6108^{+203 }_{-166}$ & $ 4.48^{+0.14 }_{-0.60}$  & $-1.96^{+0.34 }_{-0.26}$ & [0.72..0.96] & [0.63..1.54] & 1\\
3558849 & 04307 & 19 39 47.962 & +38 36 18.68 & 14.2 & $6175^{+168 }_{-194}$ & $ 4.44^{+0.07 }_{-0.27}$  & $-0.42^{+0.28 }_{-0.30}$ & [0.87..1.09] & [0.90..1.11] & 1 \\
3756801 & 01206 & 19 35 49.102 & +38 53 59.89 & 13.6 & $5796^{+162 }_{-165}$ & $ 4.12^{+0.26 }_{-0.22}$  & $-0.02^{+0.24 }_{-0.28}$ & [0.89..1.19] & [0.87..1.75]  & 2 \\
5010054$^\dagger$ &  & 19 25 59.610 & +40 10 58.40 & 14.0 & $6300^{+400 }_{-400}$ & $ 4.30^{+0.50 }_{-0.50}$  & $ 0.02^{+0.22 }_{-0.28}$ & [0.87..1.32] & [0.87..2.13] &1,2 \\
5437945 & 03791 & 19 13 53.962 & +40 39 04.90 & 13.8 & $6340^{+176 }_{-199}$ & $ 4.16^{+0.22 }_{-0.25}$  & $-0.38^{+0.28 }_{-0.30}$ & [0.90..1.24] & [0.95..1.53]  & 3\\
5522786$^\dagger$ &  & 19 13 22.440 & +40 43 52.75 &  9.3 & $8600^{+300 }_{-300}$ & $ 4.20^{+0.20 }_{-0.20}$  & $ 0.07^{+0.14 }_{-0.59}$ & [1.86..2.19] & [1.63..2.12]  &2 \\
5536555 &   & 19 30 57.482 & +40 44 10.97 &   13.5 & $5996^{+155 }_{-159}$ & $ 4.49^{+0.06 }_{-0.28}$  & $-0.48^{+0.30 }_{-0.26}$ & [0.69..0.98] & [0.67..1.38]  & 1\\
5652983 & 00371 & 19 58 42.276 & +40 51 23.36 & 12.2 & $5198^{+95  }_{ -95}$ & $ 3.61^{+0.02 }_{-0.02}$  & \nodata & [1.13..1.61] & [2.70..3.23]  & 3\\
5732155$^\dagger$ &  & 19 53 42.132 & +40 54 23.76 & 15.2 & $6000^{+400 }_{-400}$ & $ 4.20^{+0.50 }_{-0.50}$  & $-0.04^{+0.22 }_{-0.30}$ & [0.87..1.33] & [0.82..2.25]  & 2\\
5951458 &   & 19 15 57.979 & +41 13 22.91 &   12.7 & $6258^{+170 }_{-183}$ & $ 4.08^{+0.28 }_{-0.23}$  & $-0.50^{+0.30 }_{-0.30}$ & [0.77..1.19] & [0.70..2.34]  & 1\\
6191521 & 00847 & 19 08 37.032 & +41 33 56.84 & 15.2 & $5665^{+181 }_{-148}$ & $ 4.56^{+0.05 }_{-0.27}$  & $-0.58^{+0.34 }_{-0.26}$ & [0.77..0.92] & [0.75..0.88]  & 2\\
6436029 & 02828 & 19 18 09.317 & +41 53 34.15 & 15.8 & $4817^{+181 }_{-131}$ & $ 4.50^{+0.08 }_{-0.84}$  & $ 0.42^{+0.06 }_{-0.24}$ & [0.79..0.88] & [0.75..0.84]  & 3\\
7619236 & 00682 & 19 40 47.518 & +43 16 10.24 & 13.9 & $5589^{+102 }_{-108}$ & $ 4.23^{+0.13 }_{-0.12}$  & $ 0.34^{+0.10 }_{-0.14}$ & [0.93..1.12] & [0.98..1.36] & 3\\
8012732 &   & 18 58 55.079 & +43 51 51.18 &   13.9 & $6221^{+166 }_{-249}$ & $ 4.29^{+0.12 }_{-0.38}$  & $ 0.20^{+0.16 }_{-0.32}$ & [0.77..1.07] & [0.75..1.69] & 3\\
8410697 &   & 18 48 44.594 & +44 26 04.13 &   13.4 & $5918^{+157 }_{-152}$ & $ 4.37^{+0.14 }_{-0.24}$  & $-0.42^{+0.30 }_{-0.26}$ & [0.74..1.08] & [0.66..1.85]  & 1\\
8510748 &   & 19 48 19.891 & +44 30 56.12 &   11.6 & $7875^{+233 }_{-309}$ & $ 3.70^{+0.28 }_{-0.10}$  & $ 0.04^{+0.17 }_{-0.38}$ & [1.36..2.40] & [1.20..4.23]  & 1\\
8540376 &   & 18 49 30.607 & +44 41 40.52 &   14.3 & $6474^{+178 }_{-267}$ & $ 4.31^{+0.10 }_{-0.33}$  & $-0.16^{+0.23 }_{-0.32}$ & [0.84..1.23] & [0.70..1.82]  & 1,2\\
8636333 & 03349 & 19 43 47.585 & +44 45 11.23 & 15.3 & $6247^{+175 }_{-202}$ & $ 4.49^{+0.04 }_{-0.27}$  & $-0.34^{+0.26 }_{-0.30}$ & [0.86..1.03] & [0.87..1.01] & 2 \\
9214713$^\dagger$ & 00422 & 19 21 33.559 & +45 39 55.19 & 14.7 & $6200^{+400 }_{-400}$ & $ 4.40^{+0.50 }_{-0.50}$  & $-0.30^{+0.26 }_{-0.30}$ & [0.84..1.17] & [0.79..1.66] & 3 \\
9413313 &   & 19 41 40.915 & +45 54 12.56 &   14.1 & $5359^{+167 }_{-143}$ & $ 4.40^{+0.13 }_{-0.39}$  & $ 0.02^{+0.28 }_{-0.26}$ & [0.72..1.17] & [0.66..2.24]  & 2\\
9662267$^\dagger$ &  & 19 47 10.274 & +46 20 59.68 & 14.9 & $6000^{+400 }_{-400}$ & $ 4.50^{+0.50 }_{-0.50}$  & $-0.06^{+0.22 }_{-0.30}$ & [0.88..1.21] & [0.79..1.52] & 2 \\
9663113 & 00179 & 19 48 10.901 & +46 19 43.32 & 14.0 & $6065^{+155 }_{-180}$ & $ 4.42^{+0.08 }_{-0.26}$  & $-0.28^{+0.28 }_{-0.30}$ & [0.85..1.10] & [0.91..1.15]  & 2\\
9704149 &   & 19 16 39.269 & +46 25 18.48 &   15.1 & $5897^{+155 }_{-169}$ & $ 4.53^{+0.03 }_{-0.28}$  & $-0.16^{+0.24 }_{-0.30}$ & [0.73..0.99] & [0.67..1.02]  & 1\\
9838291 &   & 19 39 02.134 & +46 40 39.11 &   12.9 & $6123^{+141 }_{-177}$ & $ 4.47^{+0.05 }_{-0.29}$  & $-0.14^{+0.22 }_{-0.30}$ & [0.76..1.08] & [0.72..1.42]  & 1\\
10024862 &   & 19 47 12.602 & +46 56 04.42 &   15.9 & $6616^{+169 }_{-358}$ & $ 4.33^{+0.08 }_{-0.31}$  & $ 0.07^{+0.19 }_{-0.39}$ & [0.89..1.24] & [0.82..1.39]  & 1,3\\
10255705$^\dagger$ &  & 18 51 24.912 & +47 22 38.89 & 12.9 & $5300^{+300 }_{-300}$ & $ 3.80^{+0.40 }_{-0.40}$  & $-0.12^{+0.33 }_{-0.30}$ & [0.98..1.40] & [1.62..3.23] & 2\\
10403228$^{\dagger\dagger}$ &   & 19 24 54.410 & +47 32 59.93 &   16.1 & $3386^{+50  }_{ -50}$ & $ 4.92^{+0.06 }_{-0.07}$  & $ 0.00^{+0.10 }_{-0.10}$ & [0.27..0.37] & [0.28..0.38] &  1\\
10460629 & 01168 & 19 10 20.830 & +47 36 00.07 & 14.0 & $6449^{+163 }_{-210}$ & $ 4.23^{+0.16 }_{-0.27}$  & $-0.32^{+0.24 }_{-0.30}$ & [0.94..1.29] & [1.02..1.47]  & 2\\
10525077 & 05800 & 19 09 30.737 & +47 46 16.28 & 15.4 & $6091^{+164 }_{-213}$ & $ 4.42^{+0.06 }_{-0.30}$  & $-0.04^{+0.22 }_{-0.30}$ & [0.89..1.13] & [0.91..1.11]  & 2\\
10842718 &   & 18 47 47.285 & +48 13 21.36 &   14.6 & $5754^{+159 }_{-156}$ & $ 4.38^{+0.12 }_{-0.24}$  & $-0.06^{+0.26 }_{-0.26}$ & [0.74..1.12] & [0.65..1.90]  & 1\\
10850327 & 05833 & 19 06 21.895 & +48 13 12.97 & 13.0 & $6277^{+155 }_{-187}$ & $ 4.43^{+0.07 }_{-0.28}$  & $-0.46^{+0.28 }_{-0.30}$ & [0.87..1.10] & [0.90..1.10] & 3 \\
10960865 &   & 18 52 52.675 & +48 26 40.13 &   14.2 & $5547^{+196 }_{-154}$ & $ 4.05^{+0.34 }_{-0.26}$  & $ 0.02^{+0.26 }_{-0.26}$ & [0.73..1.19] & [0.62..2.42]  & 1\\
11465813 & 00771 & 19 46 47.666 & +49 18 59.33 & 15.2 & $5520^{+83  }_{-110}$ & $ 4.47^{+0.04 }_{-0.14}$  & $ 0.48^{+0.08 }_{-0.16}$ & [0.88..1.03] & [0.87..0.99]  & 3\\
11558724 &   & 19 26 34.094 & +49 33 14.65 &   14.7 & $6462^{+177 }_{-270}$ & $ 4.32^{+0.10 }_{-0.35}$  & $-0.08^{+0.22 }_{-0.32}$ & [0.81..1.22] & [0.70..1.80]  & 1\\
11716643 & 05929 & 19 35 27.665 & +49 48 01.04 & 14.7 & $5830^{+155 }_{-164}$ & $ 4.54^{+0.03 }_{-0.28}$  & $-0.14^{+0.24 }_{-0.28}$ & [0.79..0.93] & [0.77..0.87]  & 3\\
12066509 &   & 19 36 12.245 & +50 30 56.09 &   14.7 & $6108^{+149 }_{-192}$ & $ 4.47^{+0.04 }_{-0.30}$  & $ 0.07^{+0.15 }_{-0.33}$ & [0.80..1.11] & [0.76..1.32]  & 1\\
12356617 & 00375 & 19 24 48.286 & +51 08 39.41 & 13.3 & $5755^{+112 }_{-112}$ & $ 4.10^{+0.14 }_{-0.13}$  & $ 0.24^{+0.14 }_{-0.14}$ & [0.98..1.25] & [1.39..1.96]  & 2\\
12454613$^\dagger$ &  & 19 12 40.656 & +51 22 55.88 & 13.5 & $5500^{+280 }_{-280}$ & $ 4.60^{+0.30 }_{-0.30}$  & $ 0.00^{+0.24 }_{-0.24}$ & [0.82..1.00] & [0.77..1.00]  & 2\\

\enddata

\tablecomments{Targets with follow-up spectroscopic observations are marked with an $^\dagger$. Their stellar properties are based on MOOG analysis. We report 1-$\sigma$ range for stellar mass and radius. $^{\dagger\dagger}$: Stellar mass and radius are adopted from \citet{Huber2014}.}

\end{deluxetable}

\clearpage
\newpage

\clearpage

\begin{deluxetable}{lcccccccccccccccc}
\tabletypesize{\tiny}
 \setlength{\tabcolsep}{0.02in} 
\tablewidth{0pt}
\tablecaption{AO Sensitivity to Companions \label{tab:ao_params}}
\tablehead{
\multicolumn{7}{c}{\textbf{Kepler}} &
\multicolumn{3}{c}{\textbf{Observation}} &
\multicolumn{6}{c}{\textbf{Limiting Delta Magnitude}} \\
\colhead{\textbf{KIC}} &
\colhead{\textbf{KOI}} &
\colhead{\textbf{Kmag}} &
\colhead{\textbf{$i$}} &
\colhead{\textbf{$J$}} &
\colhead{\textbf{$H$}} &
\colhead{\textbf{$K$}} &
\colhead{\textbf{Companion}} &
\colhead{\textbf{Instrument}} &
\colhead{\textbf{Filter}} &
\colhead{\textbf{0.1}} &
\colhead{\textbf{0.2}} &
\colhead{\textbf{0.5}} &
\colhead{\textbf{1.0}} &
\colhead{\textbf{2.0}} &
\colhead{\textbf{4.0}} &
\colhead{\textbf{Orbital Solutions}} \\
\colhead{\textbf{}} &
\colhead{\textbf{}} &
\colhead{\textbf{[mag]}} &
\colhead{\textbf{[mag]}} &
\colhead{\textbf{[mag]}} &
\colhead{\textbf{[mag]}} &
\colhead{\textbf{[mag]}} &
\colhead{\textbf{within 5$^{\prime\prime}$}} &
\colhead{\textbf{}} &
\colhead{\textbf{}} &
\colhead{\textbf{[$^{\prime\prime}$]}} &
\colhead{\textbf{[$^{\prime\prime}$]}} &
\colhead{\textbf{[$^{\prime\prime}$]}} &
\colhead{\textbf{[$^{\prime\prime}$]}} &
\colhead{\textbf{[$^{\prime\prime}$]}} &
\colhead{\textbf{[$^{\prime\prime}$]}} &
\colhead{\textbf{in Table}} 
}

\startdata

2158850 &  & 10.863 & 10.726 & 9.855 & 9.570 & 9.529 & no &    NIRC2&       $K_S$&     3.8&     3.8&     5.9&     6.7&     6.7&     6.7 & 1\\
3558849 & 04307 & 14.218 & 14.035 & 13.092 & 12.819 & 12.766 & no &    NIRC2&       $K_S$&     3.3&     3.3&     5.4&     6.9&     7.0&     6.8 & 1 \\
3756801 & 01206 & 13.642 & 13.408 & 12.439 & 12.099 & 12.051 & no &    NIRC2&       $K_S$&     2.1&     4.2&     5.2&     5.3&     5.3&     5.3  & 2\\
5010054 &  & 13.961 & 13.710 & 12.797 & 12.494 & 12.412 & no &    NIRC2&       $K_S$&     2.0&     3.9&     4.9&     5.0&     4.9&     4.9  & 1,2\\
5010054 &  & 13.961 & 13.710 & 12.797 & 12.494 & 12.412 & no &  Robo-AO&       $i$&     0.2&     0.5&     2.3&     3.8&     4.6&     4.7  & 1,2\\
5437945 & 03791 & 13.771 & 13.611 & 12.666 & 12.429 & 12.367 & no &    NIRC2&       $J$&     2.4&     3.5&     5.3&     5.7&     5.7&     5.7  & 3\\
5437945 & 03791 & 13.771 & 13.611 & 12.666 & 12.429 & 12.367 & no &    NIRC2&       $K_S$&     2.8&     4.6&     6.3&     6.9&     7.0&     6.9  & 3\\
5522786 &  & 9.350 & 9.572 & 9.105 & 9.118 & 9.118 & no &    NIRC2&       $K_S$&     1.5&     4.6&     5.9&     6.5&     6.5&     6.5  & 2\\
5522786 &  & 9.350 & 9.572 & 9.105 & 9.118 & 9.118 & no &  Robo-AO&       $i$&    0.0&     0.4&     2.7&     4.6&     6.9&     8.0  & 2\\
5536555 &  & 13.465 & 13.285 & 12.313 & 11.971 & 11.933 & no &    NIRC2&       $K_S$&     3.6&     3.6&     5.3&     5.6&     5.5&     5.5  & 1\\
5652983 & 00371 & 12.193 & 11.895 & 10.723 & 10.289 & 10.169 & no &    NIRC2&       $K_S$&     2.2&     4.2&     5.5&     5.7&     5.8&     5.7  & 3\\
5732155 &  & 15.195 & 14.978 & 14.006 & 13.705 & 13.621 & yes &    NIRC2&       $K_S$&     2.1&     3.5&     4.8&     4.9&     5.0&     5.0  & 2\\
5732155 &  & 15.195 & 14.978 & 14.006 & 13.705 & 13.621 & no &  Robo-AO&       $i$&     0.6&     0.9&     2.2&     3.2&     3.4&     3.6  & 2\\
5951458 &  & 12.713 & 12.556 & 11.640 & 11.381 & 11.323 & no &    NIRC2&       $K_S$&     3.8&     3.8&     5.9&     6.7&     6.5&     6.6  & 1\\
6191521 & 00847 & 15.201 & 14.970 & 13.935 & 13.585 & 13.569 & no &    NIRC2&       $K_S$&     2.0&     3.7&     4.9&     5.1&     5.1&     5.1 & 2 \\
6436029 & 02828 & 15.768 & 15.369 & 14.041 & 13.506 & 13.429 & no &    NIRC2&       $K_S$&     1.2&     2.7&     4.4&     4.7&     4.7&     4.6  & 3\\
7619236 & 00682 & 13.916 & 13.692 & 12.688 & 12.378 & 12.260 & no &    NIRC2&       $K_S$&     1.8&     3.8&     4.8&     5.0&     5.0&     4.9  & 3\\
8012732 &  & 13.922 & 13.727 & 13.094 & 12.794 & 12.790 & no &  Robo-AO&       $i$&     0.2&     0.5&     1.8&     3.5&     4.3&     4.3  & 3\\
8410697 &  & 13.424 & 13.238 & 12.281 & 11.933 & 11.922 & no &    NIRC2&       $K_S$&     3.9&     4.6&     6.5&     7.0&     7.1&     7.1  & 1\\
8510748 &  & 11.614 & 11.638 & 10.967 & 10.898 & 10.861 & yes &    NIRC2&       $K_S$&     4.9&     4.9&     6.5&     6.8&     6.8&     6.6  & 1\\
8540376 &  & 14.294 & 14.151 & 13.259 & 13.013 & 12.965 & no &    NIRC2&       $K_S$&     4.0&     4.0&     4.3&     4.3&     4.3&     4.3  & 1,2\\
8540376 &  & 14.294 & 14.151 & 13.259 & 13.013 & 12.965 & no &  Robo-AO&       $i$&     0.2&     0.4&     1.9&     3.4&     4.1&     4.2  & 1,2\\
8636333 & 03349 & 15.292 & 15.113 & 14.192 & 13.890 & 13.880 & yes &    NIRC2&       $H$&     1.0&     2.0&     3.6&     4.8&     5.0&     5.0 & 2 \\
8636333 & 03349 & 15.292 & 15.113 & 14.192 & 13.890 & 13.880 & yes &    NIRC2&       $K_S$&     1.0&     2.3&     3.9&     4.6&     4.7&     4.6  & 2\\
9413313 &  & 14.116 & 13.835 & 12.733 & 12.335 & 12.227 & no &    NIRC2&       $K_S$&     1.9&     3.7&     4.8&     4.9&     4.9&     4.9  & 2\\
9413313 &  & 14.116 & 13.835 & 12.733 & 12.335 & 12.227 & no &  Robo-AO&       $i$&     0.4&     0.7&     2.4&     3.5&     3.9&     3.9  & 2\\
9662267 &  & 14.872 & 14.667 & 13.670 & 13.385 & 13.339 & no &    NIRC2&       $K_S$&     1.6&     3.4&     4.9&     5.0&     5.1&     5.1  & 2\\
9662267 &  & 14.872 & 14.667 & 13.670 & 13.385 & 13.339 & no &  Robo-AO&       $i$&     0.4&     0.6&     2.4&     3.2&     3.8&     3.8  & 2\\
9663113 & 00179 & 13.955 & 13.765 & 12.823 & 12.545 & 12.502 & no &    NIRC2&       $K_S$&     2.2&     3.9&     4.8&     4.9&     4.9&     4.9  & 2\\
9704149 &  & 15.102 & 14.897 & 13.896 & 13.538 & 13.454 & no &    NIRC2&       $K_S$&     1.7&     3.2&     4.4&     4.7&     4.8&     4.7  & 1\\
9704149 &  & 15.102 & 14.897 & 13.896 & 13.538 & 13.454 & no &  Robo-AO&       $i$&     0.6&     0.8&     2.1&     3.1&     3.5&     3.5  & 1\\
9838291 &  & 12.868 & 12.703 & 11.826 & 11.548 & 11.496 & no &    NIRC2&       $K_S$&     4.3&     4.3&     6.4&     7.2&     7.4&     7.2  & 1\\
10024862 &  & 15.881 & 15.712 & 14.846 & 14.551 & 14.541 & no &    NIRC2&       $K_S$&     0.7&     1.8&     2.2&     2.2&     2.2&     2.3  & 1,3\\
10024862 &  & 15.881 & 15.712 & 14.846 & 14.551 & 14.541 & no &  Robo-AO&       $i$&     0.8&     1.1&     2.0&     2.6&     2.8&     2.8  & 1,3\\
10255705 &  & 12.950 & 12.678 & 11.560 & 11.105 & 11.021 & yes &    NIRC2&       $H$&     2.5&     4.0&     5.8&     6.1&     6.3&     6.2  &2 \\
10255705 &  & 12.950 & 12.678 & 11.560 & 11.105 & 11.021 & yes &    NIRC2&       $J$&     2.0&     3.3&     5.2&     5.6&     5.7&     5.7  & 2\\
10255705 &  & 12.950 & 12.678 & 11.560 & 11.105 & 11.021 & yes &    NIRC2&       $K_S$&     2.4&     4.2&     5.6&     5.7&     5.8&     5.8  & 2\\
10255705 &  & 12.950 & 12.678 & 11.560 & 11.105 & 11.021 & yes &  Robo-AO&       $i$&     0.2&     0.3&     1.7&     3.1&     4.6&     5.0  & 2\\
10460629 & 01168 & 13.997 & 13.851 & 12.923 & 12.672 & 12.595 & no &    NIRC2&       $K_S$&     2.1&     3.7&     4.7&     4.8&     4.7&     4.7  & 2\\
10525077 & 05800 & 15.355 & 15.163 & 14.143 & 13.868 & 13.753 & no &    NIRC2&       $K_S$&     1.9&     3.4&     4.5&     4.8&     4.8&     4.8  & 2\\
10850327 & 05833 & 13.014 & 12.872 & 11.993 & 11.711 & 11.666 & no &    NIRC2&       $K_S$&     2.4&     4.1&     5.1&     5.4&     5.4&     5.3  & 3\\
11465813 & 00771 & 15.207 & 15.068 & 13.678 & 13.317 & 13.253 & yes &    NIRC2&       $H$&     1.0&     2.5&     4.7&     5.2&     5.0&     5.2  & 3\\
11465813 & 00771 & 15.207 & 15.068 & 13.678 & 13.317 & 13.253 & yes &    NIRC2&       $J$&     0.9&     2.0&     4.2&     4.6&     4.6&     4.7  & 3\\
11465813 & 00771 & 15.207 & 15.068 & 13.678 & 13.317 & 13.253 & yes &    NIRC2&       $K_S$&     1.5&     3.2&     4.1&     4.2&     4.3&     4.3  & 3\\
11716643 & 05929 & 14.692 & 14.485 & 13.483 & 13.095 & 13.092 & no &    NIRC2&       $K_S$&     2.1&     3.4&     4.2&     4.3&     4.3&     4.3  & 3\\
12356617 & 00375 & 13.293 & 13.111 & 12.137 & 11.842 & 11.791 & yes &    PHARO&       $K_S$&     0.1&     0.8&     2.5&     4.0&     4.9&     5.0 &  2\\
12454613 &  & 13.537 & 13.306 & 12.326 & 11.929 & 11.867 & no &    NIRC2&       $K_S$&     2.0&     4.2&     5.3&     5.4&     5.5&     5.5  & 2\\
12454613 &  & 13.537 & 13.306 & 12.326 & 11.929 & 11.867 & no &  Robo-AO&       $i$&     0.3&     0.4&     1.8&     3.4&     4.5&     4.7  & 2\\

\enddata

\end{deluxetable}

\clearpage
\newpage

\clearpage

\begin{deluxetable}{lcccccccccccc}
\tabletypesize{\tiny}
 \setlength{\tabcolsep}{0.02in} 
\tablewidth{0pt}
\tablecaption{AO Detections \label{tab:ao_detection}}
\tablehead{
\multicolumn{7}{c}{\textbf{Kepler}} &
\multicolumn{6}{c}{\textbf{Companion}} \\
\colhead{\textbf{KIC}} &
\colhead{\textbf{KOI}} &
\colhead{\textbf{Kmag}} &
\colhead{\textbf{$i$}} &
\colhead{\textbf{$J$}} &
\colhead{\textbf{$H$}} &
\colhead{\textbf{$K$}} &
\colhead{\textbf{sep.}} &
\colhead{\textbf{P.A.}} &
\colhead{\textbf{$\Delta i$}} &
\colhead{\textbf{$\Delta J$}} &
\colhead{\textbf{$\Delta H$}} &
\colhead{\textbf{$\Delta K$}} \\
\colhead{\textbf{}} &
\colhead{\textbf{}} &
\colhead{\textbf{[mag]}} &
\colhead{\textbf{[mag]}} &
\colhead{\textbf{[mag]}} &
\colhead{\textbf{[mag]}} &
\colhead{\textbf{[mag]}} &
\colhead{\textbf{[$^{\prime\prime}$]}} &
\colhead{\textbf{[deg]}} &
\colhead{\textbf{[mag]}} &
\colhead{\textbf{[mag]}} &
\colhead{\textbf{[mag]}} &
\colhead{\textbf{[mag]}} 
}

\startdata

5732155 &  & 15.195 & 14.978 & 14.006 & 13.705 & 13.621 & 0.93 &    221.1&       &     &     &     4.94 \\
8510748 &  & 11.614 & 11.638 & 10.967 & 10.898 & 10.861 & 0.17 & 111.9 &     &&     &     3.13\\
8636333 & 03349 & 15.292 & 15.113 & 14.192 & 13.890 & 13.880 & 0.32 &    266.2&       &     &     1.58&     1.71 \\
10255705 &  & 12.950 & 12.678 & 11.560 & 11.105 & 11.021 & 1.06 &    164.1&       1.94&     2.27&     2.37&     2.40 \\
11465813 & 00771 & 15.207 & 15.068 & 13.678 & 13.317 & 13.253 & 1.77 &    282.7 &    &   0.77&     0.74&     0.65 \\
12356617 & 00375 & 13.293 & 13.111 & 12.137 & 11.842 & 11.791 & 3.10 &    305.4&       &     &     &     4.42 \\

\enddata

\tablecomments{Typical uncertainties for companion separation (sep.), position angle (P. A.) and differential magnitude $\Delta$Mag are 0$^{\prime\prime}$.05, 0$^\circ$.5 and 0.1 mag. The uncertainties are estimated based on companion injection simulation~\citep{Wang2015}. }

\end{deluxetable}

\clearpage
\newpage

\clearpage

\begin{deluxetable}{lcccccccccc}
\tabletypesize{\tiny}
 \setlength{\tabcolsep}{0.01in} 
\tablewidth{0pt}
\tablecaption{Planet Confidence\label{tab:planet_confidence}}
\tablehead{
\colhead{\textbf{KIC}} &
\colhead{\textbf{Epoch}} &
\colhead{\textbf{Depth}} &
\colhead{\textbf{Offset}} &
\colhead{\textbf{$\delta$ Offset}} &
\colhead{\textbf{Significance}} &
\colhead{\textbf{\#Planet}} &
\colhead{\textbf{R$_{\rm{P}}$}} &
\colhead{\textbf{Confidence}} &
\colhead{\textbf{Comment}} &
\colhead{\textbf{Orbital Solutions}}	\\
\colhead{\textbf{}} &
\colhead{\textbf{BKJD}} &
\colhead{\textbf{(ppm)}} &
\colhead{\textbf{(mas)}} &
\colhead{\textbf{(mas)}}	&
\colhead{\textbf{($\sigma$)}}	&
\colhead{\textbf{Candidates}}	&
\colhead{\textbf{(R$_\oplus$)}}	&
\colhead{\textbf{}} &
\colhead{\textbf{}}  &
\colhead{\textbf{in Table}} 
}

\startdata

2158850 & 411 & 169 & 0.2 & 0.3 & 0.5 & 1 & 1.6 & 0.946 &  & 1 \\
3558849 & 279 & 3969 & 1.6 & 2.1 & 0.8 & 1 & 6.9 & 0.997 & Validated & 1 \\
3756801 & 448 & 1296 & 0.7 & 1.3 & 0.6 & 1 & 5.1 & 0.980 &  & 2 \\
5010054 & 1500 & 441 & 0.7 & 0.9 & 0.8 & 1 & 3.4 & 0.984 &  & 1 \\
5010054 & 356 & 784 & 0.3 & 0.8 & 0.4 & 1 & 4.6 & 0.883 &  & 2 \\
5437945 & 139 & 2209 & 0.3 & 0.8 & 0.4 & 2 & 6.4 & 0.999 & Validated & 3 \\
5522786 & 282 & 81 & 0.3 & 0.3 & 1.1 & 1 & 1.9 & 0.960 &  & 2 \\
5536555 & 370 & 576 & 0.7 & 1.3 & 0.5 & 1 & 2.7 & 0.922 &  & 1 \\
5536555 & 492 & 576 & 0.5 & 0.9 & 0.6 & 1 & 2.7 & 0.922 &  & 1 \\
5652983 & 244 & 12321 & 0.1 & 1.1 & 0.1 & 2 & 35.9 & \nodata & Large RV and radius & 3 \\
5732155 & 536 & 3481 & 5.0 & 4.1 & 1.2 & 1 & 9.9 & \nodata & Nearby Companion & 2 \\
5951458 & 423 & 1600 & 0.1 & 0.6 & 0.1 & 1 & 6.6 & 0.998 & Validated & 1 \\
6191521 & 382 & 4624 & 0.7 & 2.8 & 0.3 & 1 & 6.0 & 0.966 &  & 2 \\
6436029 & 458 & 2209 & 7.0 & 10.6 & 0.7 & 2 & 4.1 & 0.955 &  & 3 \\
7619236 & 185 & 5929 & 1.6 & 1.2 & 1.3 & 1 & 9.9 & 0.984 &  & 3 \\
8012732 & 391 & 5476 & 1.1 & 1.6 & 0.7 & 1 & 9.8 & 0.972 &  & 3 \\
8410697 & 542 & 5184 & 0.4 & 1.4 & 0.3 & 1 & 9.8 & 0.996 &  & 1 \\
8510748 & 1536 & 144 & 0.2 & 0.3 & 0.7 & 1 & 3.6 & \nodata & Nearby Companion & 1 \\
8540376 & 1516 & 324 & 1.0 & 1.5 & 0.7 & 3 & 2.4 & 0.999 & Validated & 1 \\
8540376 & 1520 & 900 & 0.2 & 1.6 & 0.1 & 3 & 4.1 & 0.999 & Validated & 2 \\
8636333 & 271 & 1936 & 0.9 & 2.1 & 0.4 & 2 & 4.5 & \nodata & Nearby Companion & 2 \\
9413313 & 485 & 6400 & 1.5 & 1.5 & 1.0 & 1 & 12.6 & 0.983 &  & 3 \\
9662267 & 481 & 1225 & 0.5 & 1.7 & 0.3 & 1 & 4.5 & 0.961 &  & 2 \\
9663113 & 306 & 1681 & 0.6 & 1.2 & 0.5 & 2 & 4.6 & 0.999 & Validated & 2 \\
9704149 & 419 & 2916 & 1.7 & 3.3 & 0.5 & 1 & 5.0 & 0.965 &  & 1 \\
9838291 & 582 & 1849 & 0.3 & 0.7 & 0.4 & 1 & 5.1 & 0.992 &  & 1 \\
10024862 & 878 & 9604 & 2.9 & 5.1 & 0.6 & 1 & 11.8 & 0.730 &  & 1 \\
10024862 & 359 & 2116 & 0.2 & 5.4 & 0.0 & 1 & 5.5 & 0.304 &  & 3 \\
10255705 & 545 & 1156 & 0.9 & 0.7 & 1.3 & 1 & 8.9 & \nodata & Nearby Companion & 2 \\
10460629 & 228 & 784 & 0.1 & 1.5 & 0.0 & 1 & 3.8 & 0.973 &  & 2 \\
10525077 & 335 & 2500 & 1.7 & 3.8 & 0.5 & 2 & 5.5 & 0.998 & Validated & 2 \\
10850327 & 470 & 1024 & 0.2 & 0.6 & 0.4 & 1 & 3.5 & 0.977 &  & 3 \\
11465813 & 209 & 18496 & 1.0 & 3.3 & 0.3 & 1 & 13.8 & \nodata & Nearby Companion & 3 \\
11716643 & 434 & 2209 & 1.7 & 2.8 & 0.6 & 1 & 4.2 & 0.708 &  & 3 \\
12356617 & 239 & 4761 & 0.5 & 1.2 & 0.4 & 1 & 12.5 & \nodata & Nearby Companion & 2 \\
12454613 & 490 & 1089 & 1.2 & 1.8 & 0.7 & 1 & 3.2 & 0.960 &  & 2 \\

\enddata


\end{deluxetable}

\clearpage


\end{document}